\documentclass[useAMS,usenatbib,onecolumn]{mn2e}
\usepackage{graphicx}
\usepackage{amsmath}
\usepackage{longtable,array}
\usepackage{lscape}
\usepackage{color}
\usepackage{epsfig}
\usepackage{url}

\newcommand{\ai}{{\it ab initio}}

\newcommand{\cm}{cm$^{-1}$}

\newcommand{\htwoo}{H$_2$O}

\newcommand{\hato}{H$_2$$^{16}$O}

\newcommand{\gtot}{g$_{tot}$}

\def\a0{{$a_{\rm 0}$}}

\title[ExoMol line lists  XXX: Water]{ExoMol molecular line lists XXX: a complete high-accuracy line list for water}
\author[Polyansky et al]{
Oleg L. Polyansky$^{1,2}$, Aleksandra A. Kyuberis$^{2}$, Nikolai F. Zobov$^{2}$ ,
\newauthor Jonathan Tennyson$^{1}$\thanks{Email: j.tennyson@ucl.ac.uk}, Sergei N. Yurchenko$^{1}$,
Lorenzo Lodi $^{1}$\\
$^1$Department of Physics and Astronomy, University College London, London WC1E 6BT, UK\\
$^{2}$Institute of Applied Physics, Russian Academy of Sciences,
Ulyanov Street 46, Nizhny Novgorod, Russia 603950.}

\date{\today}

\date{Accepted XXXX. Received XXXX; in original form XXXX}

\pagerange{\pageref{firstpage}--\pageref{lastpage}} \pubyear{2017}

\begin{document}

\maketitle

\label{firstpage}

\begin{abstract}
A new line list for H$_2$$^{16}$O is presented.
This line list,  which is  called POKAZATEL, includes transitions between 
rotation-vibrational energy levels up to 41~000~\cm\ in energy
and is the most complete to date.
The potential energy surface (PES) used for producing the line list
was obtained by fitting a high-quality \ai\ PES 
to experimental energy levels with energies of 41~000~\cm\ and 
for rotational excitations up to $J=5$.
The final line list comprises all energy levels up to 41~000~\cm\
and rotational angular momentum $J$ up to 72.
An accurate \ai\ dipole moment surface (DMS) was used for the calculation 
of line intensities and reproduces high-precision experimental intensity 
data with an accuracy close to 1\%.
The final line list uses empirical energy levels whenever they are available,
to ensure that line positions are reproduced as accurately as possible.
The POKAZATEL line list contains over 5 billion
transitions and is available from the ExoMol website
(\url{www.exomol.com}) and the CDS database.

\end{abstract}
\begin{keywords}
molecular data – opacity – planets and satellites: atmospheres – stars: atmospheres – 
stars: low-mass - stars: brown dwarfs.
astronomical data bases: miscellaneous.
\end{keywords}

\section{Introduction}

Water is prevalent in the Universe. In particular, the existence of
water in a wide range of hot astronomical environments has led to the
computation of very extensive line lists of rotation-vibration
transitions both for the main H$_2$$^{16}$O isotopologue
\citep{jt143,jt197,ps97,jt378} as well as for its minor isotopologues
\citep{jt469,ps97,jt438,jt665}. The most widely used water line lists
are probably the ones due to \citet{ps97}, henceforth referred to as
the Ames line list, and to \citet{jt378}, henceforth BT2.  The Ames
line list contains approximately 300 million lines while BT2 contains
500 million lines.  BT2 provided the main input for water in the 2010
release of the HITEMP database \citep{jt480}; it has since been
subject to a number of validations by comparison with laboratory
measurements \citep{14BoWeHy,15AlWeMa.H2O,16MeSaxx.H2O}.  Such
comparisons have shown that the Ames line list is often more accurate
than BT2 for transitions with wavelengths longer than 1 $\mu$m
(wavenumbers $<$ 10~000~\cm) but drops in accuracy at shorter
wavelengths.  By virtue of its greater number of lines, BT2 gives an
excellent coverage for high temperatures up to 3000 K but it is
missing significant flux at higher temperatures and for shorter
wavelengths. Thus none of these line lists can be considered to be
fully satisfactory.

The BT2 line list has been used as the basis for a number of atmospheric models
such as the widely used BT-Settl model of \citet{BT-Settl}.
However, there is increasing evidence of the presence of water
on objects hotter than 3000~K, in which case the coverage 
offered by BT2 is inadequate.
For example, water has been observed at an effective
temperature of over 4000 K in sunspots \citep{06SoWiSc.H2O} and
in a variety of giant stars with temperatures between 3500 and 5000 K
\citep{98JeSaxx.H2O,01Tsujix.H2O,06RyRiHa.H2O,12AbPaBu.H2O,15RyLaFa.H2O},
on dwarf stars with temperatures up to 4000 K \citep{14RaReAl.H2O}
and on variable stars whose atmospheres can also reach these
temperatures  \citep{jt357,08PaEvKe.H2O}.

At slightly lower temperatures, water was the first molecule
to be observed in exoplanetary atmospheres \citep{jt400} and it
is now known to be a common constituent of hot Jupiters
\citep{jt488,16IySwZe.H2O} and other exoplanets \citep{14FrDeBe.H2O}.
Some of these exoplanet observations require high accuracy
laboratory data \citep{13BiDeBr.exo,14BrDeBi.exo}.

Apart from modelling hot objects, transitions involving highly excited
energy levels may be important in environments far from thermodynamic 
equilibrium; for example, water fluorescence 
on comets can occur from very highly excited levels 
\citep{jt330,jt349,jt402}; such observations are not always
well-understood \citep{jt402}. Similarly observations with the
Atacama Large Millimeter Array (ALMA) are beginning to probe maser
emission from vibrationally excited water
\citep{12HiKiHo.H2O,16HiKiHo.H2O}; modelling water maser emission 
requires very extensive transition datasets \citep{16GrBaRi.H2O}.
Finally, extensive water line lists are also important for
many terrestrial applications, such as modelling and monitoring of
emissions from combustion engines \citep{07KrAnCa.H2O,10ReSa.H2O} or 
studying high-explosive blast waves \citep{11CaLiPi.H2O}.  

There have been a number of important developments since the
computation of the Ames and the BT2 line lists, which suggests that we now
in a position to compute an H$_2$$^{16}$O line list which is
both more comprehensive, indeed effectively complete, and more accurate
than either of these. In particular, improved theoretical methods have
led to the development of both potential energy surfaces (PES)
\citep{jt394,08LaTaTy.H2O,jt428,11BuPoZo.H2O,jt375,jt472,jt550,jt714} and 
dipole moment surfaces (DMS) \citep{08LaTaTy.H2O,jt424,jt509} with significantly
improved accuracy, and nuclear motion calculations which extend
all the way to dissociation \citep{jt230,01LiGuxx.H2O,jt472}. 
Indeed, our ability to compute high-accuracy
transition intensities is leading to such computations
\citep{jt512,jt690,jt687} replacing measurements in
standard compilations such as HITRAN \citep{jt691s}. In addition
work by an IUPAC (International Union of Pure and Applied
Chemistry) task group (Tennyson et al. 2009, 2010, 2013, 2014a, 2014b)
has led
to the determination of accurate experimental water energy levels; such very accurate,
experimentally derived energy levels can be
used to replace computed ones, resulting in both
near-perfect reproduction of laboratory transition frequencies
and in the prediction of many unobserved line positions with similar
accuracy.

In this work we exploit these advances to produce a new line list for the main 
water isotopologue H$_2$$^{16}$O. 
This work is performed as part of the ExoMol project
\citep{jt528}, which aims to provide molecular line lists for exoplanet
and other hot atmospheres. A unique feature of the resulting line list,
which we call POKAZATEL, is that it is not simply complete up to some
given temperature as is usual for ExoMol line lists of polyatomic
systems \citep{jt731}. Instead, the aim is to capture {\it all}
bound-to-bound transitions in the system, which implies considering all
energy levels lying below the dissociation limit. POKAZATEL is thus the first
complete rotation-vibrational line list for a polyatomic molecule. Besides 
covering all temperatures for which the water molecule exists, completeness
significantly extends the wavelength range of the line list. In this
context, we note the recent detection of near ultra-violet water
absorptions in the Earth's atmosphere by \citet{jt645}.

Our ability to calculate an accurate and complete water line list is
based on the five following factors: (i) the availability of spectroscopically
accurate \ai\ potential energy surface (PES) describing energies
up to the lowest dissociation pathway \citep{jt550,jt472}; (ii)
the ability to fit this \ai\ PES to empirical energy levels, which significantly
improves the accuracy of computed line positions; (iii) an efficient programme 
suite, DVR3D \citep{jt338,jt626}, which
allows us to compute accurate energies, wavefunctions, dipole moment
integrals and intensities of transitions up to dissociation; (iv) availability
of spectroscopic data covering not only the conventional infrared
and optical regions below 26 000 \cm\ \citep{jt360,jt218,jt203,jt285}
but also multiphoton spectra probing the region up to \citep{jt467} 
and even exceeding \citep{jt494} dissociation; such experimental input
allowed us to accurately characterise our new water PES up to dissociation; 
(v) progress in computer hardware, especially in terms of storage, 
allowed us to undertake comprehensive computations
previously impractical; for example, nuclear-motion wavefunctions relative to a single
value of the $J$ angular momentum often occupy more than 1 Tb of storage.

This paper is organized as follows.  In section II we describe the
PES used in this work.  Section III presents details on line
position and line intensity calculations.  In section IV we compare
our calculated energy levels with 
experimental values and with existing line lists.
Section V presents the POKAZATEL line list and discusses some ways of
using it for modelling water spectra.  Section VI concludes the paper.

\section{Calculation  of the PES}
The major, distinguishing feature of POKAZATEL with respect to its
predecessors \citep{ps97,jt197,jt378} is its completeness in both 
vibrational and rotational states. The $J=0$
vibrational energy levels included in POKAZATEL reach energies of 
40000~\cm, just below the \hato\ dissociation energy $D_0=$41~146.1~\cm\
 \citep{jt549}. To ensure completeness in rotational levels we
determined the highest $J$ for which the lowest rotational energy of
the ground vibrational state 
is below 41~000~\cm. This $J$ turns out to be $J = 72$. In other words,
for $J =73$ all the levels lie above 41~000~\cm\ and were not considered. 
The inclusion of metastable levels beyond dissociation 
is, in principle, also possible but would require a significant extension
of the methodology employed in this work. Besides, 
such transitions are likely to be important only at temperatures 
at which water is effectively decomposed and so they are not
expected to contribute significantly to the molecular opacity.

In order to calculate energy levels up to 41~000~\cm\
accurately we require two things. First, a computer program for
solving the rotation-vibrational Schr\"odinger equation 
capable of computing all the required states to the necessary
accuracy. Second, we need an accurate PES capable of reproducing to
high accuracy (to approximately 0.05~\cm) experimentally known energy levels 
and covering all energies up to dissociation.
The DVR3D program of \citet{jt338} satisfies our first requirement and has
recently 
benefitted from a number of algorithmic improvements as part of the
ExoMol project \citep{jt640,jt635,jt626} which proved vital for
completing the necessary calculations.

A water PES fulfilling our second requirement was not available, so 
we produced one as part of this work.  
Our previous, spectroscopically determined water
PESs \citep{jt182,jt308,11BuPoZo.H2O} combined with new \ai\
calculations of the \htwoo\ PES provided us with a very good
starting point for constructing our new PES.
However, even with these surfaces available, it was not easy to
produce a PES which both extends up to dissociation and provides 
near-spectroscopic accuracy. Our fitted PES is based on two separate
datasets of water energy levels. The first set consists of  conventional
spectroscopic data up to 25~000~\cm, also used in our  previously fits
\citep{jt308, 11BuPoZo.H2O}; the second set comprises energy levels from
27 000 \cm\
up to $D_0$ obtained using
two-photon and three-photon action spectroscopy \citep{07MaMuZoSh,08GrMaZoSh,jt467}.

Initially, we tried to produce a single PES reproducing the data belonging to both sets;
this proved to be impossible, as any attempt to reproduce the high-lying
energy levels to better than 1 \cm\  resulted in an unacceptable deterioration in 
low-energy levels.
Eventually we decided to follow  \citet{Varandas96} and use two separate
PES representations joined by a  switching function:

\begin{equation} \label{main_func_form}
 V_{\rm glob} = V_{\rm low}f(E) + V_{\rm up}(1-f(E))~, 
\end{equation}
\begin{equation*} 
f=0.5[1+\tanh(\gamma\Delta E)]
\end{equation*}
\begin{equation*} 
 \gamma=\gamma_0 + \gamma_1\Delta E^2
\end{equation*}
\begin{equation*} 
 \Delta E=V_u-E_0
\end{equation*}
\begin{equation} \label{main_func_f}
 \gamma_0=1/500, \gamma_1=1/500^3, E_0=35000
\end{equation}
where the values for the constants are appropriate for energies in wavenumbers.

The upper and lower surfaces employ the same functional form but have different coefficients
\begin{equation} \label{main_func_form1}
V(S_1,S_2,S_3)= V_0+\left[\sum_{i,j,k}{f_{ijk}S_1^iS_2^jS_3^k}\right]x_3+V_{\rm HH}+x_1+x_2
\end{equation}
\begin{equation*} 
 x_1=D[\exp(-2\alpha\Delta{r_1})-2\exp(-\alpha\Delta{r_1})]+D
\end{equation*}
\begin{equation*} 
 x_2=D[\exp(-2\alpha\Delta{r_2})-2\exp(-\alpha\Delta{r_2})]+D
\end{equation*}
\begin{equation*}                           
 x_3=\exp[-b_1(\Delta{r_1}^2+\Delta{r_2}^2)]
\end{equation*}
\begin{equation}                           
 V_{\rm HH}=82000\exp(-6.2r_{\rm HH})
\end{equation}
\begin{equation} \label{valence_coord}
S_1=(r_1 +r_2)/2 -r_e, S_2=(r_1 -r_2)/2, S_3 = \cos{\theta} -\cos{\theta_e}, \Delta{r_i}=r_i-r_e
\end{equation}
where units of \AA\ and \cm\ are used throughout.  $r_1$, $r_2$ and
$\theta$ are the standard bond lengths and bond angle of water, and
$r_e=0.9586$ \AA\ and $\theta_e=104.48^\circ$ are fixed to 
reference equilibrium values 
\citep{jt355}. The functions $x_1$ and $x_2$ are Morse
potentials for each of the OH bonds, $x_3$ is a damping function and
the term $V_{\rm HH}$ is a function of $r_{\rm HH}$ representing the
distance between the H atoms and is introduced to avoid artificial minima
in the region where the H atoms are close to each other
\citep{92ChLixx.H2O}.  Other non-linear constants were fixed as
follows: $b_1=2.15$~\AA$^{-1}$, $D=43900$ \cm,
$\alpha=2.2668$~\AA$^{-1}$.  The coefficients $f_{ijk}$ were determined
as discussed below.

The switch between two potentials ($E_0=35~000$ \cm) was
chosen significantly above 26~000~\cm\ to minimize the influence of the
upper PES, $V_{\rm up}$, on low-lying levels.  
The starting point
for $V_{\rm low}$ was the potential of \citet{11BuPoZo.H2O}, while the
starting point for $V_{\rm up}$ was the \ai\ PES by
\citet{jt472}. 
$V_{\rm low}$ was then determined using a fitting procedure similar
to the one employed by \citet{11BuPoZo.H2O} for the levels below 26~000~\cm;
in this region a set of 1562 empirical energy levels with $J=0,2,5$ was used.

For the fit for $V_{\rm up}$ we took as starting point the \ai\ surface \cite{jt472} and
then performed a fit of 734 levels with $J=0,2$ up to 41~000~\cm.  
Of the 41 known $J=0$  empirical levels above 27~000~\cm,
15 levels  had to be excluded from the fit.  For $J=2$ levels about 25 \%\
of the high-lying levels were excluded.
Low-energy levels below 27~000~\cm\ were also included in the fit but they were
assigned a weight 5 to 10 times lower than high-energy ones.


All fits used the approach of \citet{03YuCaJe.PH3}, in which the fitted
potential energy surface is partially constrained by a set of reference \ai\ data 
in order to prevent the emergence of nonphysical behaviour such 
as artificial peaks or troughs.
Specifically, the following functional was minimised:
\begin{equation}
\label{e:F-all}
F = \sum_i (E_{i}^{\rm (obs)}-E_{i}^{\rm (calc)})^2 w_{i}^{\rm en} +
    d \sum_k (V_{k}^{\rm (\ai)}-V_{k}^{\rm (calc)})^2 w_{k}^{\rm PES},
\end{equation}
where $E_{i}$ are the ro-vibrational energy term values, $V_k$ represents the
value of the \ai\ PES at the $k$-th geometry,
$w_{i}$ are the corresponding weight factors for the individual
energies/geometries normalized to one
and $d$ is a further factor defining the relative importance
of the \ai\ PES relative to the experimental energies. 
A total of 1460 \ai\ points were chosen for this constraint, covering 
the energy region up to about 50~000~\cm\ , which 
corresponds to O--H bond lengths and H--O--H interbond angles ranging
from 0.65 to 2.7~\AA\ and from 35$^{\circ}$ to 177$^{\circ}$, respectively.
The final PES was found to deviate from the \ai\ set of points by less
than 40 \cm\ for all geometries, which indicates that it preserves a
physically correct behaviour also for geometries uncharacterised by the 
available experimental energy levels.


The weights in Eq.~\eqref{e:F-all} are normalized
as follows:
\begin{equation}
  \sum_{i} w_i^{\rm en} +  \sum_k d w_{k}^{\rm PES}  = 1.
\end{equation}

The minimum of the function $F$ was then found using a simple 
steepest decent algorithm by simultaneously
fitting the potential parameters both to the experimental energies and to
\ai\ points \citep{03YuCaJe.PH3}. The required derivatives of the
energies with respect to the potential parameters were computed using
the Hellmann-Feyman theorem. In our analytical representation the
potential parameters are included linearly, which simplifies the
evaluation of the integrals:
\begin{equation}\label{e:Hellmann-Feynman}
\frac{\partial E_n}{\partial f_{ijk}} = \langle \psi_{n} \vert
\frac{\partial \Delta V }{\partial f_{ijk}} \vert \psi_{n} \rangle =
\langle \psi_{n} \vert \ S_1^iS_2^jS_3^kx_3~
\vert \psi_{n} \rangle .
\end{equation}
where $E_n$ and  $\psi_{n}$ are the eigenvalues and eigenfunctions of the
ro-vibrational Hamiltonian  $H$, respectively, $n$ is a running index, and
the $f_{ijk}$ are the coefficients of PES expansion.

For the fits we used the fitting wrapper \verb!DVR3D_SFit!
(simultaneous fit) written for DVR3D and also used by
\citet{11BuPoZo.H2O}. This is a Fortran 95 program which automatically
calls the necessary programs from the DVR3D suite, namely
\verb!DVR3DRJZ!, \verb!Rotlev3B!, and \verb!Xpect3!. This method
was used also to refine an H$_2$S \ai\ PES by \citet{jt640}. The wrapper
can be downloaded from CCPForge (\url{www.ccpforge.cse.rl.ac.uk})
as part of the DVR3D project.

Initially, before performing any fit, our initial, composite PES
gave observed minus calculated (obs$-$calc) residues of 
about 0.1 \cm\ for levels below 25~000 \cm.  We  then re-fitted
$V_{\rm low}$ using the lower levels while keeping $V_{\rm up}$ constant.  This
led to these levels being reproduced with a standard deviation, $\sigma$, of
0.04 \cm. After this, the upper part of the potential
$V_{\rm up}$ was fitted using the levels between 25~000 and 41~000~\cm.  
This final fit gave  $\sigma =0.13$~\cm\ for all levels and
 $\sigma=0.04$~\cm\ for the levels below 26~000~\cm.  

To improve the accuracy of the calculated energy levels at high 
angular momentum $Js$ we used
rotational nonadiabatic correction based on those of
\citet{Schwenke2001}.  Specifically, we used additional operators $J_{xx}$,
$J_{yy}$, and $J_{zz}$ to take into account the influence of
nonadiabatic effects on highly excited rotational states.  The
coefficients in front of these operators were treated as 
additional adjustable parameters and were optimized in 
calculations of the energy levels with rotational quantum 
number $J = 20$; this optimization fixed their values at 0.194, 0.194, and
0.14, respectively.

The final potential contains two sets of 246 constants and
is given in the supplementary material as a Fortran program.
Section IV below gives a
comparison between energy levels computed using the PES $V_{\rm glob}$ of
Eq.~(\ref{main_func_form}) with both experimentally derived ones and with
previous calculations.

\section{Nuclear motion calculations}
The PES described in the previous section was used to calculate
energy levels up to the energy of 40 000 \cm\ and for angular
momentum up to $J=72$. The programs
DVR3DRJZ and Rotlev3B from the DVR3D program suite \citep{jt338} were
used to perform the nuclear motion calculations. The energy
levels and corresponding wavefunctions were then used to calculate dipole
matrix elements using the program Dipole3. In turn, these matrix elements 
were then used in the program Spectra \citep{jt128} to
calculate the line positions and intensities for water transitions in
the region from 0 to 40 000 \cm. Nuclear masses were used in all
calculations.

The DVR3DRJZ calculations used Radau coordinates with 60 radial grid
points and 40 angular grid points. The radial coordinates were
represented using Morse-like oscillators with parameters
$r_e=3.0$, $D_e=0.25$, $\omega_{e}=0.007$ in atomic units;
associated Legendre polynomials were used for the angular coordinate.
Final vibrational matrices dimension of 5500 were diagonalised to give
basis functions for the full rotation-vibration calculation.  For the
rotational problem, the dimensions of final matrices fixed at
400$(J+1-p)$, where $J$ is the total angular momentum quantum
number and $p$ is the value of parity.  Our final energy levels
are converged to better
than 0.1~\cm\ at energies around 40~000 \cm, and significantly better
than this below 40~000 \cm.

\section{Results of the energy levels calculations}
To illustrate the accuracy of the energy levels associated with the
line list calculations, we compare computed energy levels with a
representative sample of experimentally derived ones for \htwoo.  
Of particular significance is the comparison of $J=0$ vibrational 
term values, as usually discrepancies between observed and
calculated (obs$-$calc) energy levels for any $J$ can be 
decomposed in a major $J$-independent vibrational contribution 
and in a much smaller $J$-dependent rotational one \citep{jt205}.
Table I gives a comparisons for all
experimentally known vibrational term values. These data are
representative of the general accuracy of all levels.

From Table I we can draw the following
conclusions. First, the POKAZATEL vibrational energy levels up to 18
000 \cm\ are more accurate than the ones of previous line
lists \citep{ps97,jt308}. Levels marked with stars were not used in
the fit of \citet{ps97} as they were experimentally unknown at that
time and show the largest differences with experiment in that line list.
In particular, un-starred levels below 16~000~\cm\ are reproduced in the 
PS line list with a root-mean-square (RMS) deviation of 0.06~\cm, while the 
deviation for starred levels is 0.41~\cm\ for levels up to 20~000~\cm\ (beyond this
energy the PS line list quickly becomes completely unreliable).
Note that the (un-starred) levels included in the fit of \citet{ps97} 
are reproduced better by both POKAZATEL and \citet{11BuPoZo.H2O} 
with a deviation of, respectively, 0.03~\cm\ and 0.01~\cm, while BT2 is 
comparable with PS with a mean average deviation for those levels of 0.08~\cm. Overall
levels below 25~000 \cm\ are best reproduced by the PES of 
\citet{11BuPoZo.H2O}, although POKAZATEL follows very closely. 

In the energy range between 18~000 and 30~000~\cm\ one finds 
discrepancies of up to 10 \cm\ for \citet{ps97} and up to 5 \cm\ 
for the BT2 line lists, which used a PES by \citet{jt308}.
The largest improvement of POKAZATEL with respect to previous line lists 
is for energies approaching 40 000 \cm, for which, 
\citet{ps97} gives discrepancies of up to 300 \cm\ while our
new line lists give about 0.1 \cm\ for levels included in the
fit and about 1 \cm\ for levels not included in the fit.
Of course, \citet{ps97} did not attempt to fit this region and
these results simply illustrate the well-known unreliability 
of extrapolation of spectroscopically determined PESs.

Another way to present the comparisons between calculated and
experimental values is standard deviations for separate $J$s. This
gives more details than overall standard deviation for all the levels
and is presented in the Table II.  

In Table II we report rms deviations for energy levels of given rotational 
angular momentum $J$ relative to the PES by \citet{11BuPoZo.H2O} and to our
new POKAZATEL one. One can see that for POKAZATEL energy levels up to
25 000 \cm\ the deviations with respect to experiment grow with $J$, 
going from about 0.04~\cm\ for $J=$0--5 to about 0.2~\cm\ for $J=$35--40.
For the levels up to 41 000 \cm, which are only known for
the $J \leq 7$, deviations are somewhat larger - up to
0.1 \cm. 

The PES by \citet{11BuPoZo.H2O}, which extends only up to 26~000~\cm,
gives deviations which are about half than the ones from POKAZATEL 
for energy levels below 25 000 \cm.

For intensity calculations wavefunctions obtained with the POKAZATEL PES
were used in all cases. However, for our final set of energy levels given in the
states file (see below) the following strategy was used. Where
available, empirical energies obtained using the MARVEL procedure by
the IUPAC task group \citep{jt539} are used. Transitions between these
levels should give line positions with experimental accuracy
even when these are yet to be observed. Unknown energy levels below
18~000 \cm\ and with $J \leq 50$ were generated using the PES of
Bubukina {\it et al.}  as these better reproduce observed line
positions in this low-energy range.  
Otherwise levels from the POKAZATEL PES are used. This
strategy gives the best available estimate for each energy level; our
data structure allows the states file to be further updated in the eventuality
that better (empirical) energies become available in the future (see, for example,
\citet{jt570}).

\begin{center}
\begin{longtable}{cccrrrrrl}
\caption{Comparison of calculated and experimentally derived \citep{jt539} vibrational term values in \cm\ for \hato\ for four different PESs: POKAZATEL (this work), 
BT2 \citep{jt378}, Bubukina \citep{11BuPoZo.H2O} and PS \citep{ps97}; levels marked with a star were not included by PS in their fit. For computed energy levels we report the difference (observed $-$ calculated). \label{t:big}}\\
\hline
\hline
$v_1$&$v_2$&$v_3$& Obs.& \multicolumn{4}{c}{obs. $-$ calc.}   \\ \cline{5-9}
     &     &     &     & POKAZATEL&   BT2     & Bubukina  &     PS   &   \\
\\                                                                    
0 & 1 & 0 &   1594.746 &  0.020    &  -0.13    & -0.016    &   -0.03  &   \\
0 & 2 & 0 &   3151.630 &  0.039    &  -0.05    & -0.004    &    0.00  &   \\
1 & 0 & 0 &   3657.053 &  0.004    &  -0.10    &  0.038    &    0.01  &   \\
0 & 0 & 1 &   3755.929 & -0.006    &   0.01    & -0.003    &   -0.03  &   \\
0 & 3 & 0 &   4666.790 &  0.041    &   0.07    &  0.000    &    0.00  &   \\
1 & 1 & 0 &   5234.976 &  0.021    &  -0.24    &  0.012    &   -0.05  &   \\
0 & 1 & 1 &   5331.267 &  0.006    &  -0.01    & -0.002    &    0.05  &   \\
0 & 4 & 0 &   6134.015 &  0.025    &   0.18    & -0.007    &   -0.02  &   \\
1 & 2 & 0 &   6775.094 & -0.013    &  -0.15    &  0.011    &   -0.01  &   \\
0 & 2 & 1 &   6871.520 &  0.011    &  -0.01    &  0.011    &    0.02  &   \\
2 & 0 & 0 &   7201.540 & -0.011    &   0.01    &  0.003    &   -0.01  &   \\
1 & 0 & 1 &   7249.817 & -0.066    &  -0.04    & -0.004    &   -0.04  &   \\
0 & 0 & 2 &   7445.056 &  0.003    &  -0.04    &  0.023    &   -0.06  &   \\
0 & 5 & 0 &   7542.372 & -0.009    &   0.16    &  0.024    &   -0.14 &  * \\
1 & 3 & 0 &   8273.976 & -0.018    &  -0.12    & -0.001    &   -0.07 &    \\
0 & 3 & 1 &   8373.851 &  0.003    &   0.00    &  0.011    &   -0.05 &    \\
2 & 1 & 0 &   8761.582 & -0.004    &  -0.12    & -0.004    &   -0.08 &    \\
1 & 1 & 1 &   8806.999 & -0.040    &  -0.10    & -0.004    &   -0.04 &    \\
0 & 6 & 0 &   8869.950 & -0.162    &  -0.32    & -0.212    &   -0.64 &  * \\
0 & 1 & 2 &   9000.136 &  0.000    &  -0.02    &  0.017    &    0.03 &    \\
0 & 4 & 1 &   9833.583 & -0.008    &   0.08    &  0.004    &   -0.05 &    \\
2 & 2 & 0 &  10284.364 & -0.025    &   0.01    &  0.000    &    0.01 &    \\
1 & 2 & 1 &  10328.729 & -0.047    &   0.03    &  0.001    &    0.06 &    \\
0 & 2 & 2 &  10521.758 & -0.017    &  -0.05    &  0.031    &   -0.01 &  * \\
3 & 0 & 0 &  10599.686 & -0.025    &   0.09    & -0.003    &    0.01 &    \\
2 & 0 & 1 &  10613.356 & -0.076    &  -0.03    & -0.003    &   -0.04 &    \\
1 & 0 & 2 &  10868.875 & -0.022    &   0.03    &  0.005    &   -0.02 &    \\
0 & 0 & 3 &  11032.404 & -0.050    &  -0.05    &  0.016    &   -0.06 &    \\
0 & 5 & 1 &  11242.776 & -0.008    &   0.16    &  0.002    &    0.00 &  * \\
2 & 3 & 0 &  11767.389 & -0.008    &  -0.09    &  0.004    &   -0.13 &  * \\
1 & 3 & 1 &  11813.207 & -0.034    &   0.00    & -0.007    &   -0.02 &    \\
0 & 3 & 2 &  12007.774 & -0.043    &  -0.14    &  0.011    &   -0.15 &  * \\
3 & 1 & 0 &  12139.315 & -0.031    &  -0.01    &  0.003    &   -0.04 &    \\
2 & 1 & 1 &  12151.254 & -0.057    &  -0.10    &  0.010    &   -0.07 &    \\
1 & 1 & 2 &  12407.662 & -0.016    &   0.00    &  0.010    &    0.01 &    \\
0 & 1 & 3 &  12565.006 & -0.043    &  -0.02    &  0.010    &    0.00 &    \\
3 & 2 & 0 &  13640.717 & -0.023    &   0.25    &  0.077    &    0.14 &  * \\
2 & 2 & 1 &  13652.655 & -0.033    &   0.21    &  0.021    &    0.19 &    \\
4 & 0 & 0 &  13828.275 & -0.001    &  -0.01    &  0.013    &    0.11 &    \\
3 & 0 & 1 &  13830.938 & -0.034    &  -0.07    &  0.001    &    0.10 &    \\
0 & 7 & 1 &  13835.373 &  0.035    &  -0.52    & -0.022    &   -0.48 &  * \\
1 & 2 & 2 &  13910.894 & -0.036    &   0.08    &  0.013    &    0.10 &    \\
0 & 2 & 3 &  14066.194 & -0.056    &   0.01    &  0.014    &   -0.02 &    \\
2 & 0 & 2 &  14221.159 & -0.024    &   0.10    &  0.004    &    0.03 &    \\
1 & 0 & 3 &  14318.813 & -0.053    &  -0.01    &  0.010    &    0.06 &    \\
1 & 5 & 1 &  14647.971 & -0.066    &  -0.17    &  0.002    &   -0.22 &  * \\
2 & 3 & 1 &  15119.028 & -0.008    &   0.02    &  0.014    &    0.00 &    \\
4 & 1 & 0 &  15344.503 & -0.012    &  -0.09    &  0.005    &    0.09 &    \\
3 & 1 & 1 &  15347.956 & -0.035    &  -0.10    &  0.001    &    0.11 &    \\
0 & 3 & 3 &  15534.709 & -0.050    &  -0.06    &  0.008    &   -0.12 &  * \\
2 & 1 & 2 &  15742.797 & -0.062    &  -0.01    & -0.021    &    0.01 &    \\
1 & 1 & 3 &  15832.766 & -0.053    &   0.00    &  0.016    &    0.09 &    \\
2 & 4 & 1 &  16546.319 & -0.010    &  -0.10    &  0.017    &  - 0.12 &  * \\
3 & 2 & 1 &  16821.631 &  0.007    &   0.39    & -0.000    &    0.54 &  * \\
4 & 2 & 0 &  16823.319 &  0.031    &   0.20    &  0.024    &    0.35 &  * \\
4 & 0 & 1 &  16898.842 &  0.000    &   0.31    &  0.001    &    0.56 &  * \\
2 & 2 & 2 &  17227.380 &  0.034    &   0.14    &  0.051    &    0.26 &  * \\
1 & 2 & 3 &  17312.551 & -0.077    &   0.25    &  0.005    &    0.31 &  * \\
3 & 0 & 2 &  17458.213 &  0.060    &   0.08    &  0.014    &    0.29 &  * \\
2 & 0 & 3 &  17495.528 & -0.038    &  -0.07    &  0.033    &    0.27 &  * \\
1 & 0 & 4 &  17748.107 & -0.032    &  -0.22    &  0.010    &    0.14 &  * \\
3 & 3 & 1 &  18265.821 & -0.031    &   0.04    & -0.008    &    0.12 &  * \\
5 & 1 & 0 &  18392.778 &  0.021    &   0.08    & -0.008    &    0.84 &  * \\
4 & 1 & 1 &  18393.315 &  0.003    &   0.10    & -0.015    &    0.84 &  * \\
1 & 3 & 3 &  18758.636 & -0.055    &   0.04    &  0.018    &    0.17 &  * \\
2 & 1 & 3 &  18989.960 & -0.038    &  -0.06    &  0.019    &    0.30 &  * \\
5 & 0 & 1 &  19781.103 &  0.009    &  -0.20    &  0.002    &    1.06 &  * \\
6 & 0 & 0 &  19781.323 &  0.028    &  -0.69    &  0.013    &    0.73 &  * \\
4 & 2 & 1 &  19865.285 &  0.011    &   0.74    &  0.014    &    1.79 &  * \\
2 & 2 & 3 &  20442.777 & -0.011    &   0.29    &  0.031    &         &    \\
3 & 0 & 3 &  20543.129 & -0.024    &  -0.14    &  0.034    &         &    \\
5 & 1 & 1 &  21221.827 & -0.038    &  -0.52    &  0.024    &         &    \\
4 & 3 & 1 &  21314.448 &  0.040    &   1.14    &  0.055    &         &    \\
7 & 0 & 0 &  22529.295 & -0.054    &   0.06    & -0.005    &    2.51 &  * \\
6 & 0 & 1 &  22529.441 & -0.054    &   0.08    &  0.004    &    2.50 &  * \\
7 & 0 & 1 &  25120.278 & -0.067    &   0.52    &  0.015    &    4.25 &  * \\
5 & 3 & 2 &  27502.660 & -0.232    &  -0.85    &           &         &    \\
9 & 0 & 0 &  27540.690 &  0.148    &   1.11    &           &    4.23 &  * \\
6 & 1 & 2 &  27574.910 & -0.069    &   0.50    &           &         &    \\
9 & 1 & 0 &  28934.140 &  0.833    &   3.20    &           &         &    \\
10& 0 & 0 &  29810.850 &  0.417    &           &           &    7.30 &  * \\
8 & 0 & 2 &  31071.570 & -0.034    &           &           &         &    \\
10& 1 & 0 &  31207.090 & -1.724    &           &           &         &    \\
11& 0 & 0 &  31909.679 &  0.673    &           &           &    9.14 &  * \\
11& 1 & 0 &  33144.709 & -0.080    &           &           &         &    \\
11& 0 & 1 &  33835.222 &  0.164    &           &           &   15.33 &  * \\
12& 0 & 0 &  33835.249 &  0.193    &           &           &   15.36 &  * \\
13& 0 & 0 &  35585.957 &  0.407    &           &           &   44.65 &  * \\
12& 0 & 1 &  35586.007 & -0.565    &           &           &   43.14 &  * \\
12& 2 & 0 &  36179.317 & -5.567    &           &           &          &   \\
13& 1 & 0 &  36684.047 & -2.843    &           &           &          &   \\
12& 1 & 1 &  36684.877 & -2.052    &           &           &          &   \\
9 & 1 & 3 &  36739.777 & -2.312    &           &           &          &   \\
10& 1 & 2 &  36740.597 & -1.841    &           &           &          &   \\
14& 0 & 0 &  37122.697 &  0.517    &           &           &   108.47& *  \\
13& 0 & 1 &  37122.717 &  0.556    &           &           &   108.50& *  \\
11& 3 & 1 &  37309.847 &  0.221    &           &           &         &    \\
12& 3 & 0 &  37311.277 & -0.593    &           &           &         &    \\
13& 2 & 0 &  37765.647 &  0.186    &           &           &         &    \\
14& 1 & 0 &  38153.247 &  0.029    &           &           &         &    \\
13& 1 & 1 &  38153.307 &  0.045    &           &           &         &    \\
15& 0 & 0 &  38462.517 &  1.094    &           &           &   283.17& *  \\
14& 0 & 1 &  38462.537 &  1.112    &           &           &   283.11& *  \\
14& 2 & 0 &  39123.767 &  0.240    &           &           &          &   \\
14& 1 & 1 &  39390.217 &  1.220    &           &           &          &   \\
15& 1 & 0 &  39390.257 &  1.258    &           &           &          &   \\
15& 0 & 1 &  39574.537 &  0.249    &           &           &          &   \\
16& 0 & 0 &  39574.547 &  0.218    &           &           &          &   \\
12& 1 & 2 &  40044.567 & -3.005    &           &           &          &   \\
11& 1 & 3 &  40044.671 &  2.611    &           &           &          &   \\
14& 2 & 1 &  40226.261 & -4.185    &           &           &          &   \\
13& 3 & 1 &  40262.001 & -5.579    &           &           &          &   \\
16& 1 & 0 &  40370.547 & -0.921    &           &           &          &   \\
15& 1 & 1 &  40370.781 & -0.764    &           &           &          &   \\
16& 0 & 1 &  40437.211 &  1.941    &           &           &          &   \\
12& 0 & 3 &  40704.156 & -2.329    &           &           &          &   \\
17& 0 & 1 &  40945.693 &  0.270    &           &           &          &   \\
18& 0 & 1 &  41100.053 &  1.468    &           &           &          &   \\
19& 0 & 0 &  41101.337 & -0.554    &           &           &          &   \\
15& 2 & 1 &  41121.606 &  4.564    &           &           &          &   \\
\hline\hline
\end{longtable}
\end{center}


\begin{table}
\caption{Comparison of obs$-$calc root-mean-square deviations ($\sigma$) in \cm\ as a function of rotational ($J$) states for calculations using our new POKAZATEL PES and using the PES by \citet{11BuPoZo.H2O}. The reported rms deviations are relative to energy levels up to $E_\mathrm{max}$. $N_J$ is the number of levels considered in each case.}
\begin{tabular}{rrrrrrr}
\hline
\hline
  PES=  &    \multicolumn{2}{c}{POKAZATEL}  &  \multicolumn{2}{c}{Bubukina} & \multicolumn{2}{c}{POKAZATEL}  \\
$E_\mathrm{max}$ &  \multicolumn{2}{c}{26~000~\cm} & \multicolumn{2}{c}{25~000~\cm}  & \multicolumn{2}{c}{41~000~\cm} \\
   $J$ &  $N_J$   &   $\sigma$     &  $N_J$  & $\sigma$    &  $N_J$ & $\sigma$  \\
\hline
     0 &   77   &   0.034  &   78 & 0.023&   86  &  0.113  \\
     1 &  274   &   0.039  &  270 & 0.022&  306  &  0.112  \\
     2 &  482   &   0.040  &  475 & 0.032&  548  &  0.112  \\
     3 &  679   &   0.037  &  673 & 0.029&  719  &  0.081  \\
     4 &  841   &   0.039  &  842 & 0.033&  875  &  0.065  \\
     5 &  994   &   0.047  &  997 & 0.031& 1025  &  0.075  \\
     6 & 1064   &   0.055  & 1069 & 0.032& 1074  &  0.064  \\
     7 & 1110   &   0.070  & 1104 & 0.035& 1112  &  0.072  \\
     8 & 1035   &   0.087  & 1037 & 0.037&       &         \\
     9 &  953   &   0.109  &  950 & 0.039&       &         \\
    10 &  857   &   0.125  &  851 & 0.044&       &         \\
    11 &  750   &   0.134  &  752 & 0.054&       &         \\
    12 &  676   &   0.150  &  672 & 0.063&       &         \\
    13 &  615   &   0.157  &  605 & 0.069&       &         \\
    14 &  575   &   0.169  &  560 & 0.075&       &         \\
    15 &  527   &   0.179  &  519 & 0.082&       &         \\
    16 &  494   &   0.183  &  503 & 0.097&       &         \\
    17 &  474   &   0.193  &  480 & 0.108&       &         \\
    18 &  450   &   0.194  &  460 & 0.116&       &         \\
    19 &  442   &   0.206  &  450 & 0.124&       &         \\
    20 &  432   &   0.210  &  420 & 0.126&       &         \\
    21 &  399   &   0.203  &  396 & 0.131&       &         \\
    22 &  366   &   0.209  &  358 & 0.134&       &         \\
    23 &  344   &   0.212  &  329 & 0.138&       &         \\
    24 &  316   &   0.219  &  306 & 0.140&       &         \\
    25 &  283   &   0.207  &  268 & 0.142&       &         \\
    26 &  251   &   0.193  &  234 & 0.154&       &         \\
    27 &  238   &   0.189  &  228 & 0.162&       &         \\
    28 &  218   &   0.185  &  193 & 0.163&       &         \\
    29 &  181   &   0.194  &  172 & 0.163&       &         \\
    30 &  143   &   0.190  &  124 & 0.165&       &         \\
    31 &  110   &   0.198  &      &      &       &          \\
    32 &   81   &   0.185  &      &      &       &          \\
    33 &   46   &   0.213  &      &      &       &          \\
    34 &   24   &   0.223  &      &      &       &          \\
    35 &   19   &   0.198  &      &      &       &          \\
    36 &   17   &   0.217  &      &      &       &          \\
    37 &   11   &   0.199  &      &      &       &          \\
    38 &   10   &   0.170  &      &      &       &          \\
    39 &    4   &   0.186  &      &      &       &          \\

\hline
\hline
\end{tabular}

\label{t:Jbreakdown}
\end{table}

Our nuclear-motion calculations assign to energy levels only exact quantum
numbers, namely $J$, parity and the ortho/para symmetry label. 
Furthermore, energy levels within a given $J$-parity-symmetry subset
are labelled in increasing order of energy with a counting index $i$.

However, it is convenient and standard practice to label energy levels
with approximate (normal mode) vibrational $\nu_1 \nu_2 \nu_3$ and (rigid rotor)
rotational $J K_a K_c$ quantum numbers. Assigning such labels to
every level up to dissociation is difficult \citep{jt472}
and probably formally impossible \citep{jt234}.  
Nevertheless, many energy levels can indeed be successfully labelled and
such labelling is useful in some applications, e.g. when 
considering pressure broadening.  For low $v$ and $J$ the labelling
procedure is straightforward while for higher excitations a variety of
methods were used, which will be discussed elsewhere.  In our line list
vibrational $(v_1 v_2 v_3)$ and rotational $(J_{K_a K_c})$ labels 
were assigned to more than 72~000  \htwoo\ energy levels with 
energies up to 20~000 \cm\ and $J\leq 28$.

\section{Calculation and representation of the line list}
The program suite DVR3D calculates the bound rotation-vibration energy levels and
the corresponding wavefunctions on a three dimensional grid.
Using these wavefunctions and the LTP2011S DMS  \citep{jt509} we computed
the Einstein A coefficients, $A_{if}$, for 
transitions up to $J=72$ and the energies up to 40~000 \cm.

\begin{table}

\caption{Extracts from the final states file for the POKAZATEL line list showing
portions with full quantum number assignments (upper part) and only rigorous quantum
numbers given (lower part).}

\begin{tabular}{rrcccccccc}

\hline\hline

$i$ &  $\tilde{E}$  & \gtot  & $J$ & $K_a$ & $K_c$ & $\nu_1$ & $\nu_2$ & $\nu_3$ & $S$ \\

\hline

           2 &  1594.746306  &    1  &     0 &  0 &  0 &  0 &  1 &  0 &  A1 \\
           3 &  3151.629850  &    1  &     0 &  0 &  0 &  0 &  2 &  0 &  A1 \\
           4 &  3657.053255  &    1  &     0 &  0 &  0 &  1 &  0 &  0 &  A1 \\
           5 &  4666.790461  &    1  &     0 &  0 &  0 &  0 &  3 &  0 &  A1 \\
           6 &  5234.975555  &    1  &     0 &  0 &  0 &  1 &  1 &  0 &  A1 \\
           7 &  6134.015008  &    1  &     0 &  0 &  0 &  0 &  4 &  0 &  A1 \\
           8 &  6775.093508  &    1  &     0 &  0 &  0 &  1 &  2 &  0 &  A1 \\
           9 &  7201.539855  &    1  &     0 &  0 &  0 &  2 &  0 &  0 &  A1 \\
          10 &  7445.056211  &    1  &     0 &  0 &  0 &  0 &  0 &  2 &  A1 \\
          11 &  7542.372492  &    1  &     0 &  0 &  0 &  0 &  5 &  0 &  A1 \\
          12 &  8273.975695  &    1  &     0 &  0 &  0 &  1 &  3 &  0 &  A1 \\
          13 &  8761.581581  &    1  &     0 &  0 &  0 &  2 &  1 &  0 &  A1 \\
          14 &  8869.950054  &    1  &     0 &  0 &  0 &  0 &  6 &  0 &  A1 \\
          15 &  9000.136035  &    1  &     0 &  0 &  0 &  0 &  1 &  2 &  A1 \\
          16 &  9724.179914  &    1  &     0 &  0 &  0 &  1 &  4 &  0 &  A1 \\
          17 & 10085.961796  &    1  &     0 &  0 &  0 &  0 &  7 &  0 &  A1 \\
          18 & 10284.364368  &    1  &     0 &  0 &  0 &  2 &  2 &  0 &  A1 \\
          19 & 10521.757715  &    1  &     0 &  0 &  0 &  0 &  2 &  2 &  A1 \\
          20 & 10599.685969  &    1  &     0 &  0 &  0 &  3 &  0 &  0 &  A1 \\
\\
         100 & 21703.511719  &    1   &    0 & -2 & -2 & -2 & -2 & -2 &  A1 \\
         101 & 21764.097656  &    1   &    0 & -2 & -2 & -2 & -2 & -2 &  A1 \\
         102 & 21844.693359  &    1   &    0 & -2 & -2 & -2 & -2 & -2 &  A1 \\
         103 & 21916.152344  &    1   &    0 & -2 & -2 & -2 & -2 & -2 &  A1 \\
         104 & 21972.789062  &    1   &    0 & -2 & -2 & -2 & -2 & -2 &  A1 \\
         105 & 22006.955078  &    1   &    0 & -2 & -2 & -2 & -2 & -2  & A1 \\
         106 & 22127.925781  &    1   &    0 & -2 & -2 & -2 & -2 & -2  & A1 \\
         107 & 22166.060547  &    1   &    0 & -2 & -2 & -2 & -2 & -2  & A1 \\
         108 & 22326.316406  &    1   &    0 & -2 & -2 & -2 & -2 & -2  & A1 \\
         109 & 22376.539062  &    1   &    0 & -2 & -2 & -2 & -2 & -2  & A1 \\
         110 & 22385.830078  &    1   &    0 & -2 & -2 & -2 & -2 & -2  & A1 \\
\hline\hline
\end{tabular}
\label{t:states}
\mbox{}\\

{\flushleft
$i$:   State counting number.     \\
$\tilde{E}$: State energy in \cm. \\
\gtot: Total state degeneracy.\\
$J$: Total angular momentum.            \\
$K_a$: Projection of the angular momentum in the prolate symmetric top limit. \\
$K_c$: Projection of the angular momentum in the oblate  symmetric top limit. \\
$\nu_1$:   Symmetric stretch quantum number. \\
$\nu_2$:   Bending quantum number. \\
$\nu_3$:   Asymmetric stretch quantum number. \\
$S$: State symmetry in C$_{2v}$.}
\end{table}

\begin{table}
\caption{Extract from the transitions file for POKAZATEL line list.}
\begin{tabular}{rrr}
\hline\hline
$f$ & $i$ & $A_{fi}$  \\
\hline
      596233  &     571007 & 1.9373e-02 \\
      725029  &     732339 & 1.0832e-02 \\
      329530  &     297534 & 8.0899e-04 \\
      790239  &     794617 & 1.8347e-02 \\
      221420  &     214352 & 2.3773e-02 \\
      421277  &     402418 & 6.1962e-02 \\
      438351  &     418788 & 1.2862e-01 \\
      472230  &     500166 & 2.6986e-04 \\
      442671  &     459574 & 1.7567e-03 \\
      208210  &     178893 & 3.1161e-02 \\
      380584  &     398311 & 5.4103e-03 \\
      437709  &     442656 & 4.0364e-03 \\
      623411  &     618141 & 1.3136e-02 \\
       41424  &      44438 & 6.1976e-04 \\
      638780  &     616418 & 2.2414e-04 \\
      478821  &     448210 & 7.2435e-03 \\
       92899  &      71149 & 1.3900e-03 \\
      190855  &     172844 & 3.2447e-02 \\
      429814  &     398308 & 4.1444e-03 \\
       78888  &     100775 & 3.7271e-05 \\
      735537  &     742327 & 2.4587e-04 \\
\hline\hline
\end{tabular}
\label{t:trans}
\mbox{}\\
{$f$}: Upper state counting number.  \\
{$i$}: Lower state counting number. \\
$A_{fi}$: Einstein-A coefficient in s$^{-1}$.\\
\end{table}

The ExoMol database uses a condensed format which separates transitions
into a states file (which includes quantum labels where available) and
a transitions file \citep{jt548}. Extracts from these two files are given in
Tables ~\ref{t:states} and \ref{t:trans}, respectively. 
These files, which contain 810~269 states and 5~745~071~340 transitions
can be obtained from
\url{ftp://cdsarc.u-strasbg.fr/pub/cats/J/MNRAS/xxx/yy}, or
\url{http://cdsarc.u-strasbg.fr/viz-bin/qcat?J/MNRAS//xxx/yy} as well
as the ExoMol website, \url{www.exomol.com}.

\begin{table}
\caption{Comparison of partition functions: VT \citep{jt263},  
definitive \citep{jt661}, BT2 \citep{jt378} and Ames \citep{ps97}.}
\begin{tabular}{cccccc}
\hline
\hline
     T(K) &  VT   &   POKAZATEL     & definitive  &  BT2 &  Ames  \\
\hline

100 &   35.153  &   35.15320               &    35.153 12   &35.15451   &   35.15279        \\
300 &   178.122 &   178.1210               &    178.120 6   &178.1279   &   178.1175        \\
500 &   386.333 &   386.3309               &    386.330 0   &386.3446   &   386.3224        \\
800 &   823.791 &   823.7822               &    823.780 1   &823.8028   &   823.7627        \\
1000    &   1218.319    &   1 218.276           &   1218.273    &1 218.299  &   1 218.247       \\
1200    &   1717.126    &   1 717.092       &   1717.087    &1 717.114  &   1 717.052       \\
1500    &   2713.816    &   2 713.061       &   2713.052    &2 713.078  &   2 713.002       \\
1800    &   4093.180    &   4 091.037   &   4091.024    &4 091.046  &   4 090.949       \\
2000    &   5279.984    &   5 276.323       &   5276.309    &5 276.322  &   5 276.186       \\
2500    &   9465.976    &   9 456.146   &   9456.14 &   9 455.912   &   9 455.016       \\
3000    &   15981.08    &   15 961.28       &   15961.3 &   15 956.96   &   15 949.21   \\
3200    &   19433.68    &   19 408.59       &   19408.7 &   19 397.33   &   19 380.95   \\
3400    &   23467.56    &   23 436.15       &   23436.4 &   23 409.67   &   23 377.37   \\
3500    &   25725.04    &   25 690.04       &   25690.4 &   25 650.77   &   25 606.48   \\
3600    &   28155.76    &   28 116.74       &   28117.2 &   28 059.71   &   27 999.90   \\
3800    &   33577.48    &   33 529.18       &   33530.1 &   33 415.36   &   33 310.73   \\
4000    &   39818.16    &   39 758.33       &   39760   &   39 545.59   &   39 371.57   \\
4200    &   46969.6 &   46 895.09       &   46899   &   46 519.39   &   46 242.48   \\
4400    &   55128.8 &   55 036.08       &   55045   &   54 404.58   &   53 980.91   \\
4500    &   59618.4 &   59 514.84       &   59527   &   58 709.55   &   58 192.42   \\
4600    &   64399.2 &   64 283.29       &   64300   &   63 266.73   &   62 640.65   \\ 
4800    &   74888.4 &   74 743.42       &   74775   &   73 168.15   &   72 271.03   \\
5000    &   86709.2 &   86 527.25       &   86584   &   84 166.99   &   82 916.31   \\
5200    &   99976.4 &   99 748.69       &   99847   &   96 316.52   &   94 615.14   \\
5400    &   114808.8    &   114 523.9       &   114687  &   109 664.5   &   107 400.3   \\
5500    &   122848.8    &   122 530.8       &   122739  &   116 801.3   &   114 208.9   \\
5600    &   131324.8    &   130 970.1       &   131234  &   124 252.9   &   121 298.7   \\
5800    &   149644.8    &   149 204.9       &   149619  &   140 117.5   &   136 330.8   \\
6000    &   169887.2    &   169 344.6       &   169977  &   157 287.6   &   152 511.6   \\

\hline
\hline
\end{tabular}\label{PQ}
\end{table}

In order to further improve the accuracy of the line positions presented
in the POKAZATEL line list, we produced two additional sets of energy
levels, presented as two different states files in ExoMol format,
which are made available in the supplementary material.  In the first
additional file we substituted the POKAZATEL energy levels 
up to 20 000 \cm\ with the ones calculated with the PES by 
\citep{11BuPoZo.H2O}. A comparison between the POKAZATEL PES and
the PES by \citep{11BuPoZo.H2O} is reported in Tables \ref{t:big}
and \ref{t:Jbreakdown} and shows that energy levels produced 
by \citep{11BuPoZo.H2O} are somewhat more accurate then the ones resulting
from POKAZATEL for energies up to about 20~000~\cm.

A second set of energy 
levels was produced exclusively from experimentally derived ones. 
This set of levels is significantly more limited but 
much more accurate, as its accuracy corresponds 
to the one of experimental observations. 
 
Using our calculations we also computed the partition function
for \hato\ for wide range of temperatures.
The partition function of water is important for a variety of applications
and high accuracy studies are available concentrating solely on this
quantity \citep{jt263,jt661}. Table~\ref{PQ} compares our partition
function with those from various previous studies. 
All partition functions are computed
using the HITRAN convention \citep{jt692}, adopted by the ExoMol project,
which explicitly includes the spin degeneracy of all particles.
As the value of the partition function at a given temperature always increases
when more energy levels are included in its calculation, such value can be used
as a measure of the completeness of a given line list at that
temperature \citep{jt181}.

Table~\ref{PQ} shows that the POKAZATEL partition function gives
excellent agreement with the recent \lq definitive\rq\ partition
function of \citet{jt661}.  The agreement between these and the older
partition function of \citet{jt263} is also excellent; Vidler \&\
Tennyson also considered all states up to dissociation with $J \leq
72$ but used a rather crude model for  the high-lying energies.
This illustrates an important point: for accurate partition sums at
high temperatures completeness of the energy level list is more
important than accuracy of individual levels.  Conversely, both the
BT2 and the Ames partition sums are too low at high temperatures, which
reflects the incompleteness of these line lists.
Our partition function is given in the supplementary data on a grid of 1 K.

\begin{figure}
\begin{center}
\includegraphics[width=0.55\textwidth]{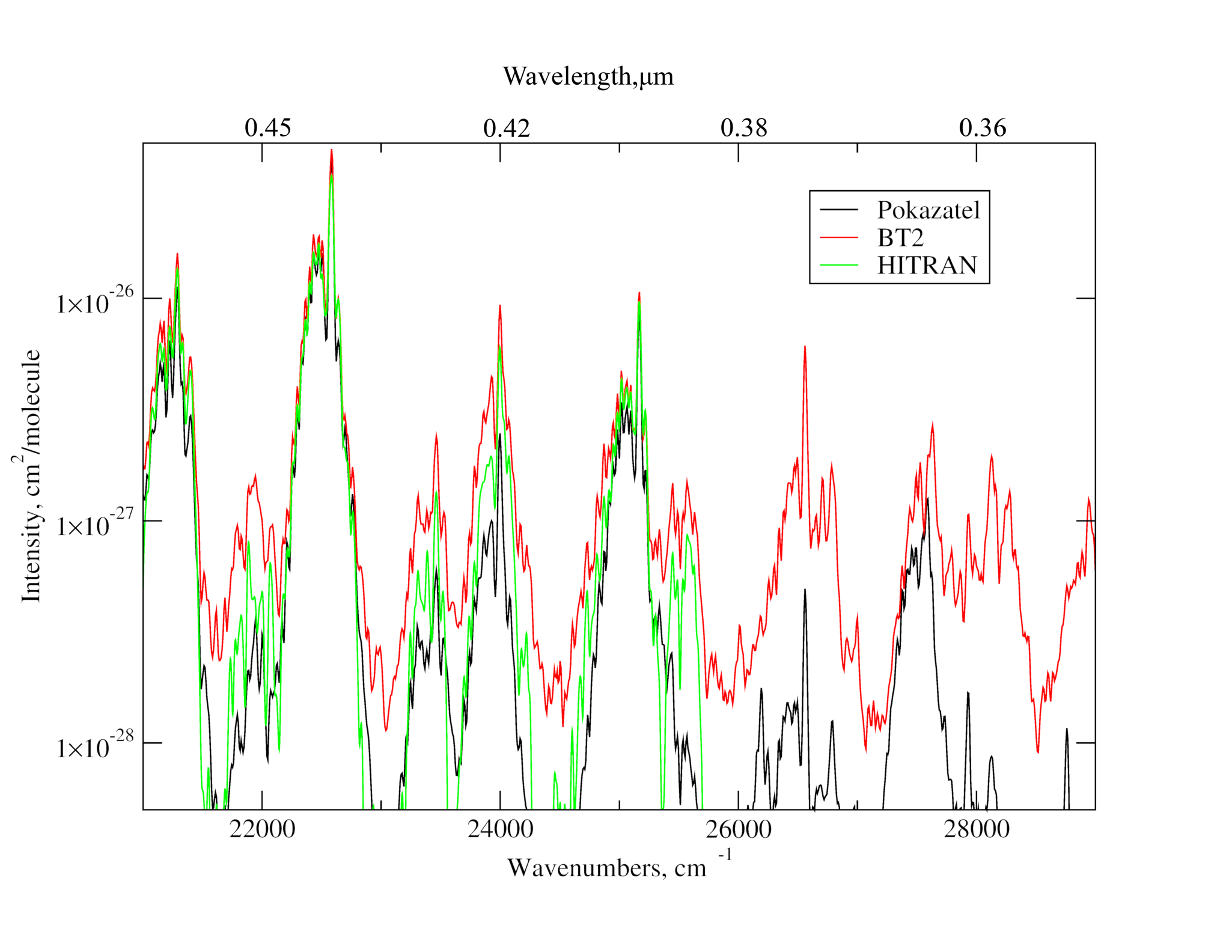}
\includegraphics[width=0.55\textwidth]{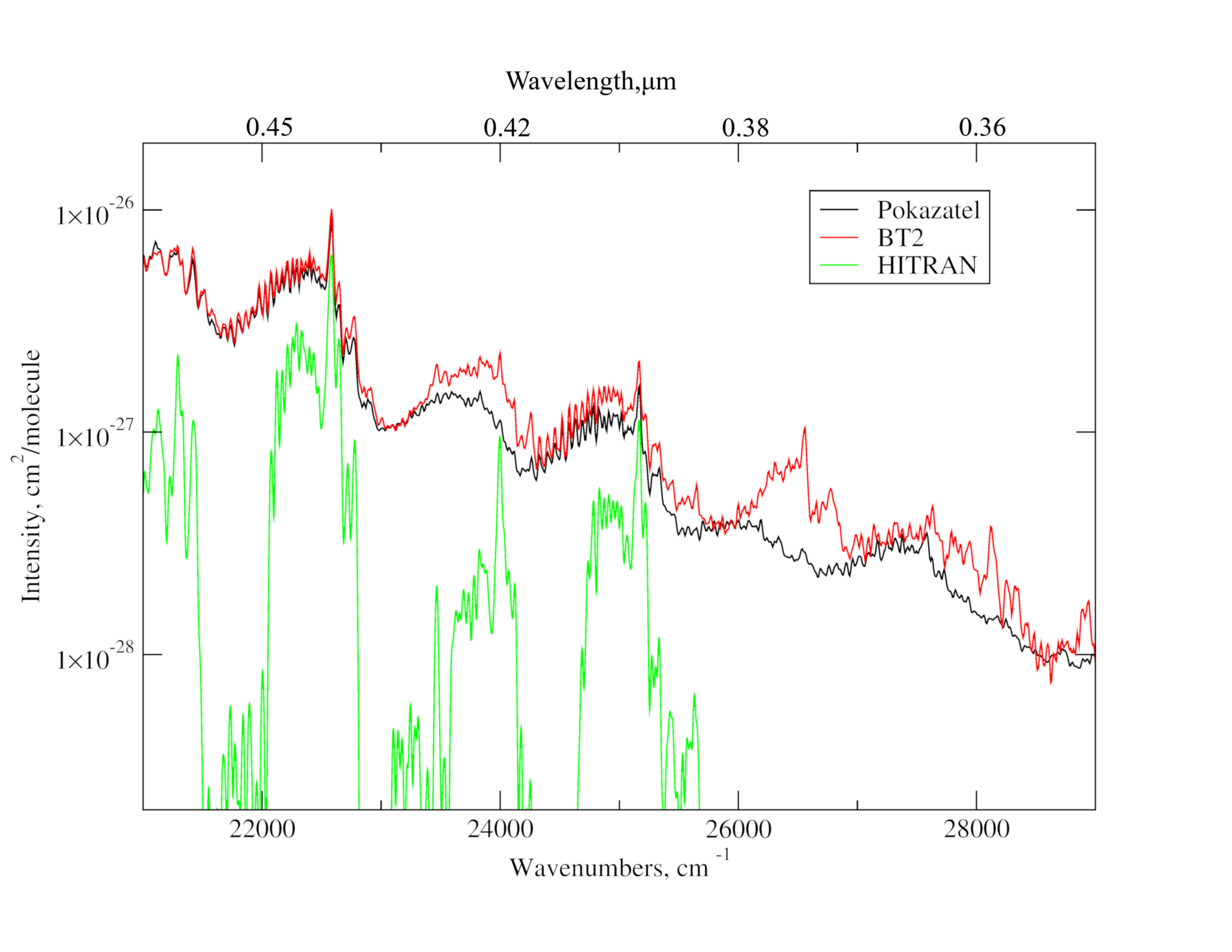}
\end{center}
\caption{\label{fig:combt2uv} Comparison of BT2 \citep{jt378},
  POKAZATEL and HITRAN 2016 \citep{jt691s} at blue and
  near-ultraviolet wavelengths.  Upper plot is for room temperature ($T$= 296
  K), lower  plot is for $T=2000$~K.}
\end{figure}

\begin{figure}
\begin{center}
\includegraphics[width=0.8\textwidth]{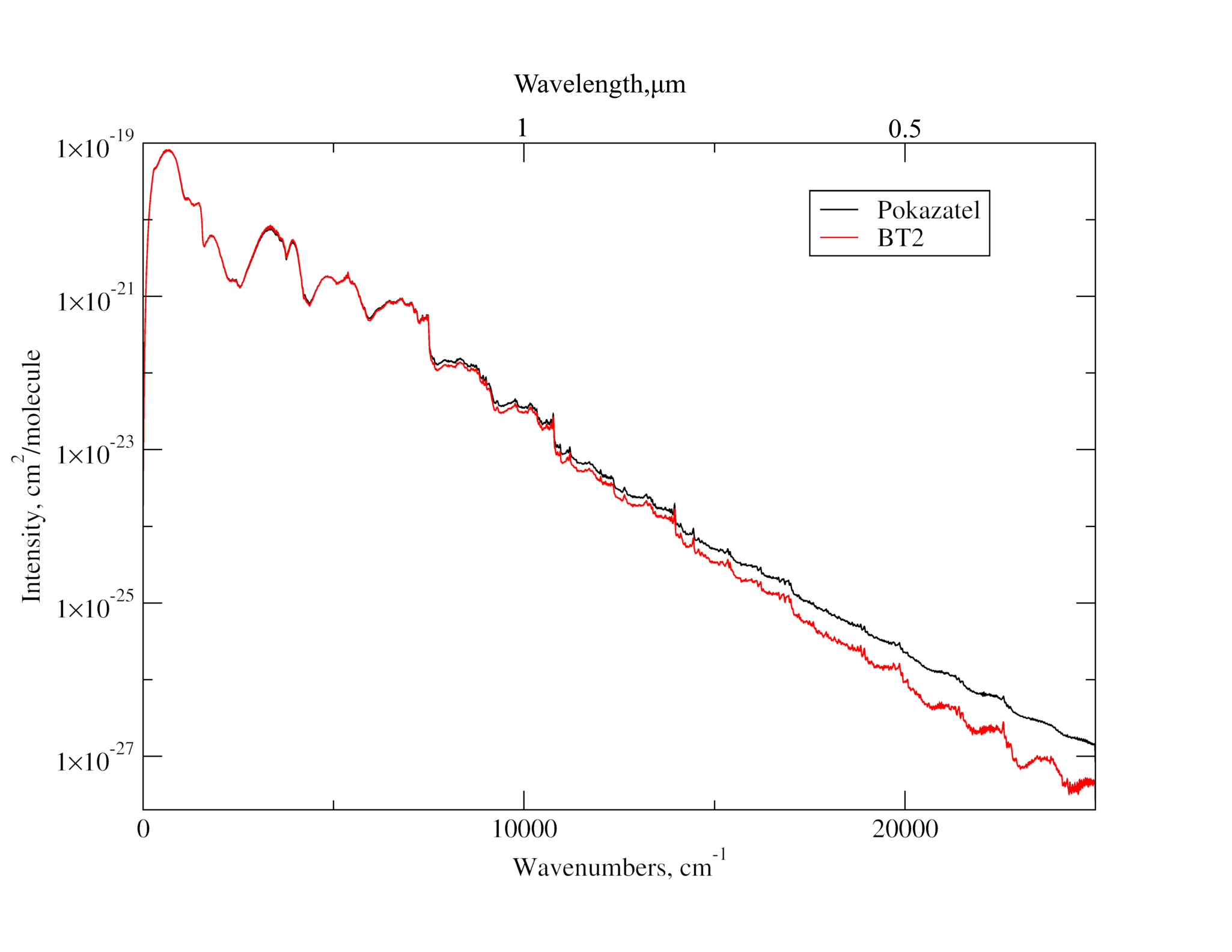}
\end{center}
\caption{\label{fig:combt4000} Comparison of absorption by
\hato\ predicted by  BT2 and POKAZATEL at $T=4000$ K; note the increased
absorption with POKAZATEL and flattened structures.}
\end{figure}

To illustrate our results Fig. \ref{fig:combt2uv} and \ref{fig:combt4000} present
plots of \htwoo\ spectra in
various spectral regions and for various temperatures.  
Below about 2500 K, at the low resolution of the plots
POKAZATEL coincides quite closely with BT2 for wavenumbers up to 25 000 \cm.

BT2 uses an energy limit of 30~000 \cm\ and was designed to
be complete for transitions below 20~000 \cm and, as a result, its
predictions are very different from the ones by POKAZATEL
in the near-UV region above 25~000~\cm, see fig.~\ref{fig:combt2uv}. 
In particular, the BT2 line list predicts a much larger
absorption in the near-UV region than POKAZATEL. 
The quality of the POKAZATEL line list in this region has been demonstrated 
in recent analysis of ultraviolet terrestrial atmospheric absorption \citep{jt645},
so we can conclusively say that BT2 overestimates absorption in this region.

At room temperature and for visible wavelengths POKAZATEL and the recent release of HITRAN \citep{jt691s} give
reasonable agreement, but HITRAN contains no data on near-ultraviolet transitions. At high temperatures,
as expected, HITRAN significantly underestimates the absorption. We note that in this region
HITEMP \citep{jt480} corresponds to BT2.

Figure \ref{fig:combt4000} shows a comparison between BT2 and
POKAZATEL for the high temperature of 4000~K. It can be seen that the
absorption spectrum predicted by BT2 is more structured
than that of POKAZATEL. The flattening of the spectrum is
characteristic of a more complete treatment, including high $J$ states
and vibrational hot bands. High temperature ($T > 3000$~K) models
relying on BT2 are therefore missing significant opacity.

\begin{figure}
\begin{center}
\includegraphics[width=0.7\textwidth]{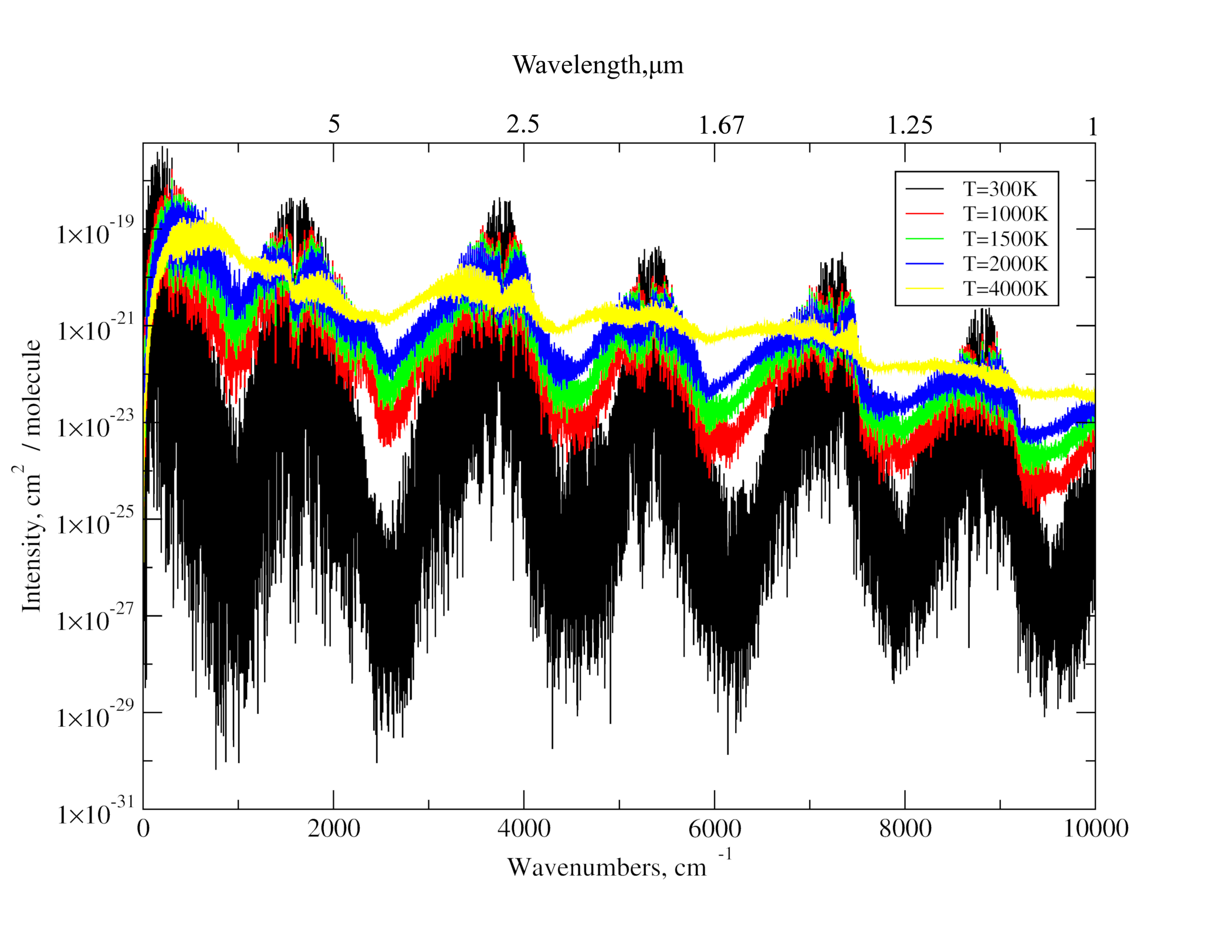}
\end{center}
\caption{\label{fig:cs}
Temperature dependence of the absorption cross-sections of the POKAZATEL line list computed. Cross sections become
increasingly flattened with increasing temperature.}
\end{figure}

Figure ~\ref{fig:cs} presents cross sections computed using ExoCross
\citep{jt708} for different temperatures using the POKAZATEL
line list. This illustrates the change in the absorption spectra with
increasing temperature.

\section{Conclusion}

We present in this work a new, very complete water line list, 
which we call POKAZATEL. The line list includes vibrational
and rotational energies up to 40 000 \cm\ and a maximum rotational angular
momentum $J=72$.  Our calculations are based on a newly developed PES for 
water which extends all to way up to the lowest-energy dissociation
pathway. The accuracy of computed energy levels is about 0.1 \cm\ for all 
the energies up to dissociation.  In the lower energy region up to 25~000~\cm\ accuracy
is better.  The line list comprises nearly 6 billion lines, an order of
magnitude more than any previous line list, and can be used for the
modelling of very hot water spectra up to the ultraviolet region.

For infrared frequencies and temperatures up to 2000 K the
overall absorption modelled by POKAZATEL is very similar to the one
by BT2 \citep{jt378}, although the accuracy of individual
lines is significantly improved as illustrated by Tables 1 and 2.
In particular, we note that recent, independent laboratory studies at
room temperature \citep{17CaMiVa.H2O,18KaStCaDa} and in flames \citep{jt712}
have strongly endorsed the accuracy of the POKAZATEL predictions.
At short wavelengths and higher
temperatures the completeness of the present line list results in 
significant opacity differences from BT2. 

An important aspect of comprehensive line lists is the treatment 
of pressure broadening. The ExoMol project
recently developed a pressure-broadening diet \citep{jt684} aimed
at including the effect of broadening by H$_2$ and He at high temperatures.
Particular attention has been paid to the broadening of water spectra
\citep{jt544,jt669}. We note that the implementation used within the
ExoMol project allows for the treatment of pressure broadening to be 
transferred between line lists and hence also to the POKAZATEL line list.
All data, including pressure broadening ones, can be found in the ExoMol
database \citep{jt631}.

\section{Acknowledgement}

This work was supported by the ERC Advanced Investigator Project 267219, the UK Natural Environment
Research Council and the Russian Fund for Fundamental Studies (grant 18-02-0057).

\bibliographystyle{mn2e}

\begin{thebibliography}{}

\bibitem[\protect\citeauthoryear{Abia, Palmerini, Busso \& Cristallo}{Abia
  et~al.}{2012}]{12AbPaBu.H2O}
Abia C.,  Palmerini S.,  Busso M.,    Cristallo S.,  {2012}, Astron.
  Astrophys., {548}, A55

\bibitem[\protect\citeauthoryear{Alberti, Weber, Mancini, Fateev \&
  Clausen}{Alberti et~al.}{2015}]{15AlWeMa.H2O}
Alberti M.,  Weber R.,  Mancini M.,  Fateev A.,    Clausen S.,  2015, J. Quant.
  Spectrosc. Radiat. Transf., 157, 14

\bibitem[\protect\citeauthoryear{{Allard}}{{Allard}}{2014}]{BT-Settl}
{Allard} F.,  2014, in {Booth} M.,  {Matthews} B.~C.,   {Graham} J.~R.,  eds,
  IAU Symposium Vol.~299 of IAU Symposium, {The BT-Settl Model Atmospheres for
  Stars, Brown Dwarfs and Planets}.
pp 271--272

\bibitem[\protect\citeauthoryear{Allard, Hauschildt, Miller \& Tennyson}{Allard
  et~al.}{1994}]{jt143}
Allard F.,  Hauschildt P.~H.,  Miller S.,    Tennyson J.,  1994, Astrophys. J.,
  426, L39

\bibitem[\protect\citeauthoryear{Azzam, Yurchenko, Tennyson \& Naumenko}{Azzam
  et~al.}{2016}]{jt640}
Azzam A. A.~A.,  Yurchenko S.~N.,  Tennyson J.,    Naumenko O.~V.,  2016, Mon.
  Not. R. Astron. Soc., 460, 4063

\bibitem[\protect\citeauthoryear{Banerjee, Barber, Ashok \& Tennyson}{Banerjee
  et~al.}{2005}]{jt357}
Banerjee D. P.~K.,  Barber R.~J.,  Ashok N.~K.,    Tennyson J.,  2005,
  Astrophys. J., 627, L141

\bibitem[\protect\citeauthoryear{Barber, Miller, Stallard, Tennyson, Hirst,
  Carroll \& Adamson}{Barber et~al.}{2007}]{jt402}
Barber R.~J.,  Miller S.,  Stallard T.,  Tennyson J.,  Hirst P.,  Carroll T.,
   Adamson A.,  2007, Icarus, 187, 167

\bibitem[\protect\citeauthoryear{Barber, Strange, Hill, Polyansky, Mellau,
  Yurchenko \& Tennyson}{Barber et~al.}{2014}]{jt570}
Barber R.~J.,  Strange J.~K.,  Hill C.,  Polyansky O.~L.,  Mellau G.~C.,
  Yurchenko S.~N.,    Tennyson J.,  2014, Mon. Not. R. Astron. Soc., 437, 1828

\bibitem[\protect\citeauthoryear{Barber, Tennyson, Harris \& Tolchenov}{Barber
  et~al.}{2006}]{jt378}
Barber R.~J.,  Tennyson J.,  Harris G.~J.,    Tolchenov R.~N.,  2006, Mon. Not.
  R. Astron. Soc., 368, 1087

\bibitem[\protect\citeauthoryear{Barletta, Shirin, Zobov, Polyansky, Tennyson,
  Valeev \& {Cs\'asz\'ar}}{Barletta et~al.}{2006}]{jt394}
Barletta P.,  Shirin S.~V.,  Zobov N.~F.,  Polyansky O.~L.,  Tennyson J.,
  Valeev E.~F.,    {Cs\'asz\'ar} A.~G.,  2006, J. Chem. Phys., 125, 204307

\bibitem[\protect\citeauthoryear{Barton, Hill, Czurylo, Li, Hyslop, Yurchenko
  \& Tennyson}{Barton et~al.}{2017}]{jt684}
Barton E.~J.,  Hill C.,  Czurylo M.,  Li H.-Y.,  Hyslop A.,  Yurchenko S.~N.,
   Tennyson J.,  2017, J. Quant. Spectrosc. Radiat. Transf., 203, 490

\bibitem[\protect\citeauthoryear{Barton, Hill, Yurchenko, Tennyson, Dudaryonok
  \& Lavrentieva}{Barton et~al.}{2017}]{jt669}
Barton E.~J.,  Hill C.,  Yurchenko S.~N.,  Tennyson J.,  Dudaryonok A.,
  Lavrentieva N.~N.,  2017, J. Quant. Spectrosc. Radiat. Transf., 187, 453

\bibitem[\protect\citeauthoryear{Beaulieu, Kipping, Batista, Tinetti, Ribas,
  Noriega-Crespo, Griffith, Campanella, Dong, Tennyson, Barber, Deroo, Fossey,
  Liang, Swain \& Yung}{Beaulieu et~al.}{2010}]{jt488}
Beaulieu J.~P.,  Kipping D.~M.,  Batista V.,  Tinetti G.,  Ribas I.,
  Noriega-Crespo S. C. J.~A.,  Griffith C.~A.,  Campanella G.,  Dong S.,
  Tennyson J.,  Barber R.~J.,  Deroo P.,  Fossey S.~J.,  Liang D.,  Swain
  M.~R.,    Yung Y.,  2010, Mon. Not. R. Astron. Soc., 409, 963

\bibitem[\protect\citeauthoryear{Birk, Wagner, Loos, Lodi, Polyansky, Kyuberis,
  Zobov \& Tennyson}{Birk et~al.}{2017}]{jt687}
Birk M.,  Wagner G.,  Loos J.,  Lodi L.,  Polyansky O.~L.,  Kyuberis A.~A.,
  Zobov N.~F.,    Tennyson J.,  2017, J. Quant. Spectrosc. Radiat. Transf.,
  203, 88

\bibitem[\protect\citeauthoryear{Birkby, de Kok, Brogi, de Mooij, Schwarz,
  Albrecht \& Snellen}{Birkby et~al.}{2013}]{13BiDeBr.exo}
Birkby J.~L.,  de Kok R.~J.,  Brogi M.,  de Mooij E. J.~W.,  Schwarz H.,
  Albrecht S.,    Snellen I. A.~G.,  {2013}, Mon. Not. R. Astron. Soc., {436},
  L35

\bibitem[\protect\citeauthoryear{Bordbar, Wecel \& Hyppanen}{Bordbar
  et~al.}{2014}]{14BoWeHy}
Bordbar M.~H.,  Wecel G.,    Hyppanen T.,  {2014}, Cumbust. Flame, {161}, 2435

\bibitem[\protect\citeauthoryear{Boyarkin, Koshelev, Aseev, Maksyutenko, Rizzo,
  Zobov, Lodi, Tennyson \& Polyansky}{Boyarkin et~al.}{2013}]{jt549}
Boyarkin O.~V.,  Koshelev M.~A.,  Aseev O.,  Maksyutenko P.,  Rizzo T.~R.,
  Zobov N.~F.,  Lodi L.,  Tennyson J.,    Polyansky O.~L.,  2013, Chem. Phys.
  Lett., 568-569, 14

\bibitem[\protect\citeauthoryear{Brogi, de Kok, Birkby, Schwarz \&
  Snellen}{Brogi et~al.}{2014}]{14BrDeBi.exo}
Brogi M.,  de Kok R.~J.,  Birkby J.~L.,  Schwarz H.,    Snellen I. A.~G.,
  {2014}, Astron. Astrophys., {565}, A124

\bibitem[\protect\citeauthoryear{Bubukina, Polyansky, Zobov \&
  Yurchenko}{Bubukina et~al.}{2011}]{11BuPoZo.H2O}
Bubukina I.~I.,  Polyansky O.~L.,  Zobov N.~F.,    Yurchenko S.~N.,  2011,
  Optics Spectrosc., 110, 160

\bibitem[\protect\citeauthoryear{Campargue, Mikhailenko, Vasilchenko, Reynaud,
  Beguier, Cermak, Mondelain, Kassi \& Romanini}{Campargue
  et~al.}{2017}]{17CaMiVa.H2O}
Campargue A.,  Mikhailenko S.~N.,  Vasilchenko S.,  Reynaud C.,  Beguier S.,
  Cermak P.,  Mondelain D.,  Kassi S.,    Romanini D.,  {2017}, J. Quant.
  Spectrosc. Radiat. Transf., {189}, 407

\bibitem[\protect\citeauthoryear{Carney, Lightstone, Piecuch \& Koch}{Carney
  et~al.}{2011}]{11CaLiPi.H2O}
Carney J.~R.,  Lightstone J.~M.,  Piecuch S.,    Koch J.~D.,  {2011}, Measure.
  Sci. Tech., {22}, {045601}

\bibitem[\protect\citeauthoryear{Child, Weston \& Tennyson}{Child
  et~al.}{1999}]{jt234}
Child M.~S.,  Weston T.,    Tennyson J.,  1999, Mol. Phys., 96, 371

\bibitem[\protect\citeauthoryear{Choi \& Light}{Choi \&
  Light}{1992}]{92ChLixx.H2O}
Choi S.~E.,  Light J.~C.,  1992, J. Chem. Phys., 97, 7031

\bibitem[\protect\citeauthoryear{{Cs\'asz\'ar}, Czako, Furtenbacher, Tennyson,
  Szalay, Shirin, Zobov \& Polyansky}{{Cs\'asz\'ar} et~al.}{2005}]{jt355}
{Cs\'asz\'ar} A.~G.,  Czako G.,  Furtenbacher T.,  Tennyson J.,  Szalay V.,
  Shirin S.~V.,  Zobov N.~F.,    Polyansky O.~L.,  2005, J. Chem. Phys., 122,
  214305

\bibitem[\protect\citeauthoryear{Cs\'asz\'ar, M\'atyus, Szidarovszky, Lodi,
  Zobov, Shirin, Polyansky \& Tennyson}{Cs\'asz\'ar et~al.}{2010}]{jt472}
Cs\'asz\'ar A.~G.,  M\'atyus E.,  Szidarovszky T.,  Lodi L.,  Zobov N.~F.,
  Shirin S.~V.,  Polyansky O.~L.,    Tennyson J.,  2010, J. Quant. Spectrosc.
  Radiat. Transf., 111, 1043

\bibitem[\protect\citeauthoryear{{Dello Russo}, Bonev, DiSanti, Gibb, Mumma,
  Magee-Sauer, Barber \& Tennyson}{{Dello Russo} et~al.}{2005}]{jt349}
{Dello Russo} N.,  Bonev B.~P.,  DiSanti M.~A.,  Gibb E.~L.,  Mumma M.~J.,
  Magee-Sauer K.,  Barber R.~J.,    Tennyson J.,  2005, Astrophys. J., 621, 537

\bibitem[\protect\citeauthoryear{{Dello Russo}, DiSanti, Magee-Sauer, Gibb,
  Mumma, Barber \& Tennyson}{{Dello Russo} et~al.}{2004}]{jt330}
{Dello Russo} N.,  DiSanti M.~A.,  Magee-Sauer K.,  Gibb E.~L.,  Mumma M.~J.,
  Barber R.~J.,    Tennyson J.,  2004, Icarus, 168, 186

\bibitem[\protect\citeauthoryear{Faure, Wiesenfeld, Tennyson \& Drouin}{Faure
  et~al.}{2013}]{jt544}
Faure A.,  Wiesenfeld L.,  Tennyson J.,    Drouin B.~J.,  2013, J. Quant.
  Spectrosc. Radiat. Transf., 116, 79

\bibitem[\protect\citeauthoryear{Fraine, Deming, Benneke, Knutson, Jordan,
  Espinoza, Madhusudhan, Wilkins \& Todorov}{Fraine
  et~al.}{2014}]{14FrDeBe.H2O}
Fraine J.,  Deming D.,  Benneke B.,  Knutson H.,  Jordan A.,  Espinoza N.,
  Madhusudhan N.,  Wilkins A.,    Todorov K.,  {2014}, Nature, {513}, 526+

\bibitem[\protect\citeauthoryear{Furtenbacher, Szidarovszky, Hruby, Kyuberis,
  Zobov, Polyansky, Tennyson \& Cs\'asz\'ar}{Furtenbacher et~al.}{2016}]{jt661}
Furtenbacher T.,  Szidarovszky T.,  Hruby J.,  Kyuberis A.~A.,  Zobov N.~F.,
  Polyansky O.~L.,  Tennyson J.,    Cs\'asz\'ar A.~G.,  2016, J. Phys. Chem.
  Ref. Data, 45, 043104

\bibitem[\protect\citeauthoryear{Gamache, Roller, Lopes, Gordon, Rothman,
  Polyansky, Zobov, Kyuberis, Tennyson, Yurchenko, Cs\'asz\'ar, Furtenbacher,
  Huang, Schwenke, Lee, Drouin, Tashkun, Perevalov \& Kochanov}{Gamache
  et~al.}{2017}]{jt692}
Gamache R.~R.,  Roller C.,  Lopes E.,  Gordon I.~E.,  Rothman L.~S.,  Polyansky
  O.~L.,  Zobov N.~F.,  Kyuberis A.~A.,  Tennyson J.,  Yurchenko S.~N.,
  Cs\'asz\'ar A.~G.,  Furtenbacher T.,  Huang X.,  Schwenke D.~W.,  Lee T.~J.,
  Drouin B.~J.,  Tashkun S.~A.,  Perevalov V.~I.,    Kochanov R.~V.,  2017, J.
  Quant. Spectrosc. Radiat. Transf., 203, 70

\bibitem[\protect\citeauthoryear{Gordon \& {et al.}}{Gordon \& {et
  al.}}{2017}]{jt691s}
Gordon I.~E.,  {et al.} 2017, J. Quant. Spectrosc. Radiat. Transf., 203, 3

\bibitem[\protect\citeauthoryear{Gray, Baudry, Richards, Humphreys, Sobolev \&
  Yates}{Gray et~al.}{2016}]{16GrBaRi.H2O}
Gray M.~D.,  Baudry A.,  Richards A. M.~S.,  Humphreys E. M.~L.,  Sobolev
  A.~M.,    Yates J.~A.,  {2016}, Mon. Not. R. Astron. Soc., {456}, 374

\bibitem[\protect\citeauthoryear{Grechko, Boyarkin, Rizzo, Maksyutenko, Zobov,
  Shirin, Lodi, Tennyson, {Cs\'asz\'ar} \& Polyansky}{Grechko
  et~al.}{2009}]{jt467}
Grechko M.,  Boyarkin O.~V.,  Rizzo T.~R.,  Maksyutenko P.,  Zobov N.~F.,
  Shirin S.,  Lodi L.,  Tennyson J.,  {Cs\'asz\'ar} A.~G.,    Polyansky O.~L.,
  2009, J. Chem. Phys., 131, 221105

\bibitem[\protect\citeauthoryear{Grechko, Maksyutenko, Zobov, Shirin,
  Polyansky, Rizzo \& Boyarkin}{Grechko et~al.}{2008}]{08GrMaZoSh}
Grechko M.,  Maksyutenko P.,  Zobov N.~F.,  Shirin S.~V.,  Polyansky O.~L.,
  Rizzo T.~R.,    Boyarkin O.~V.,  2008, J. Phys. Chem. A, 112, 10539

\bibitem[\protect\citeauthoryear{Hirota, Kim \& Honma}{Hirota
  et~al.}{2012}]{12HiKiHo.H2O}
Hirota T.,  Kim M.~K.,    Honma M.,  {2012}, Astrophys. J. Lett., {757}, L1

\bibitem[\protect\citeauthoryear{Hirota, Kim \& Honma}{Hirota
  et~al.}{2016}]{16HiKiHo.H2O}
Hirota T.,  Kim M.~K.,    Honma M.,  {2016}, Astrophys. J., {817}, 168

\bibitem[\protect\citeauthoryear{Iyer, Swain, Zellem, Line, Roudier, Rocha \&
  Livingston}{Iyer et~al.}{2016}]{16IySwZe.H2O}
Iyer A.~R.,  Swain M.~R.,  Zellem R.~T.,  Line M.~R.,  Roudier G.,  Rocha G.,
   Livingston J.~H.,  {2016}, Astrophys. J., {823}, 109

\bibitem[\protect\citeauthoryear{Jennings \& Sada}{Jennings \&
  Sada}{1998}]{98JeSaxx.H2O}
Jennings D.~E.,  Sada P.~V.,  {1998}, Science, {279}, 844

\bibitem[\protect\citeauthoryear{Kassi, Stoltmann, Casado, Daeron \&
  Campargue}{Kassi et~al.}{2018}]{18KaStCaDa}
Kassi S.,  Stoltmann T.,  Casado M.,  Daeron M.,    Campargue A.,  {2018}, J.
  Chem. Phys., {148}, 054201

\bibitem[\protect\citeauthoryear{Kranendonk, An, Caswell, Herold, Sanders,
  Huber, Fujimoto, Okura \& Urata}{Kranendonk et~al.}{2007}]{07KrAnCa.H2O}
Kranendonk L.~A.,  An X.,  Caswell A.~W.,  Herold R.~E.,  Sanders S.~T.,  Huber
  R.,  Fujimoto J.~G.,  Okura Y.,    Urata Y.,  {2007}, Optics Express, {15},
  15115

\bibitem[\protect\citeauthoryear{Kyuberis, Zobov, Naumenko, Voronin, Polyansky,
  Lodi, Liu, Hu \& Tennyson}{Kyuberis et~al.}{2017}]{jt690}
Kyuberis A.~A.,  Zobov N.~F.,  Naumenko O.~V.,  Voronin B.~A.,  Polyansky
  O.~L.,  Lodi L.,  Liu A.,  Hu S.-M.,    Tennyson J.,  2017, J. Quant.
  Spectrosc. Radiat. Transf., 203, 175

\bibitem[\protect\citeauthoryear{Lamouroux, Tashkun \& Tyuterev}{Lamouroux
  et~al.}{2008}]{08LaTaTy.H2O}
Lamouroux J.,  Tashkun S.~A.,    Tyuterev V.~G.,  2008, Chem. Phys. Lett., 452,
  225

\bibitem[\protect\citeauthoryear{Lampel, P\"{o}hler, Polyansky, Kyuberis,
  Zobov, Tennyson, Lodi, Frie{\ss}, Wang, Beirle, Platt \& Wagner}{Lampel
  et~al.}{2017}]{jt645}
Lampel J.,  P\"{o}hler D.,  Polyansky O.~L.,  Kyuberis A.~A.,  Zobov N.~F.,
  Tennyson J.,  Lodi L.,  Frie{\ss} U.,  Wang Y.,  Beirle S.,  Platt U.,
  Wagner T.,  2017, Atmos. Chem. Phys., 17, 1271

\bibitem[\protect\citeauthoryear{Li \& Guo}{Li \& Guo}{2001}]{01LiGuxx.H2O}
Li G.,  Guo H.,  {2001}, J. Mol. Spectrosc., {210}, 90

\bibitem[\protect\citeauthoryear{Lodi \& Tennyson}{Lodi \&
  Tennyson}{2008}]{jt428}
Lodi L.,  Tennyson J.,  2008, J. Quant. Spectrosc. Radiat. Transf., 109, 1219

\bibitem[\protect\citeauthoryear{Lodi, Tennyson \& Polyansky}{Lodi
  et~al.}{2011}]{jt509}
Lodi L.,  Tennyson J.,    Polyansky O.~L.,  2011, J. Chem. Phys., 135, 034113

\bibitem[\protect\citeauthoryear{Lodi, Tolchenov, Tennyson, Lynas-Gray, Shirin,
  Zobov, Polyansky, {Cs\'asz\'ar}, {van Stralen} \& Visscher}{Lodi
  et~al.}{2008}]{jt424}
Lodi L.,  Tolchenov R.~N.,  Tennyson J.,  Lynas-Gray A.~E.,  Shirin S.~V.,
  Zobov N.~F.,  Polyansky O.~L.,  {Cs\'asz\'ar} A.~G.,  {van Stralen} J.,
  Visscher L.,  2008, J. Chem. Phys., 128, 044304

\bibitem[\protect\citeauthoryear{Maksyutenko, Muenter, Zobov, Shirin,
  Polyansky, Rizzo \& Boyarkin}{Maksyutenko et~al.}{2007}]{07MaMuZoSh}
Maksyutenko P.,  Muenter J.~S.,  Zobov N.~F.,  Shirin S.~V.,  Polyansky O.~L.,
  Rizzo T.~R.,    Boyarkin O.~V.,  2007, J. Chem. Phys., 126, 241101

\bibitem[\protect\citeauthoryear{Melin \& Sanders}{Melin \&
  Sanders}{2016}]{16MeSaxx.H2O}
Melin S.~T.,  Sanders S.~T.,  {2016}, J. Quant. Spectrosc. Radiat. Transf.,
  {180}, 184

\bibitem[\protect\citeauthoryear{Mizus, Kyuberis, Zobov, Makhnev, Polyansky \&
  Tennyson}{Mizus et~al.}{2018}]{jt714}
Mizus I.~I.,  Kyuberis A.~A.,  Zobov N.~F.,  Makhnev V.~Y.,  Polyansky O.~L.,
   Tennyson J.,  2018, Phil. Trans. Royal Soc. London A, 376, 20170149

\bibitem[\protect\citeauthoryear{Mussa \& Tennyson}{Mussa \&
  Tennyson}{1998}]{jt230}
Mussa H.~Y.,  Tennyson J.,  1998, J. Chem. Phys., 109, 10885

\bibitem[\protect\citeauthoryear{Neale, Miller \& Tennyson}{Neale
  et~al.}{1996}]{jt181}
Neale L.,  Miller S.,    Tennyson J.,  1996, Astrophys. J., 464, 516

\bibitem[\protect\citeauthoryear{Partridge \& Schwenke}{Partridge \&
  Schwenke}{1997}]{ps97}
Partridge H.,  Schwenke D.~W.,  1997, J. Chem. Phys., 106, 4618

\bibitem[\protect\citeauthoryear{Pavanello, Adamowicz, Alijah, Zobov, Mizus,
  Polyansky, Tennyson, Szidarovszky, Cs\'asz\'ar, Berg, Petrignani \&
  Wolf}{Pavanello et~al.}{2012}]{jt512}
Pavanello M.,  Adamowicz L.,  Alijah A.,  Zobov N.~F.,  Mizus I.~I.,  Polyansky
  O.~L.,  Tennyson J.,  Szidarovszky T.,  Cs\'asz\'ar A.~G.,  Berg M.,
  Petrignani A.,    Wolf A.,  2012, Phys. Rev. Lett., 108, 023002

\bibitem[\protect\citeauthoryear{Pavlenko, Evans, Kerr, Yakovina, Woodward,
  Lynch, Rudy, Pearson \& Russell}{Pavlenko et~al.}{2008}]{08PaEvKe.H2O}
Pavlenko Y.~V.,  Evans A.,  Kerr T.,  Yakovina L.,  Woodward C.~E.,  Lynch D.,
  Rudy R.,  Pearson R.~L.,    Russell R.~W.,  {2008}, Astron. Astrophys.,
  {485}, 541

\bibitem[\protect\citeauthoryear{Polyansky, Jensen \& Tennyson}{Polyansky
  et~al.}{1996}]{jt182}
Polyansky O.~L.,  Jensen P.,    Tennyson J.,  1996, J. Chem. Phys., 105, 6490

\bibitem[\protect\citeauthoryear{Polyansky, Kyuberis, Lodi, Tennyson,
  Ovsyannikov \& Zobov}{Polyansky et~al.}{2017}]{jt665}
Polyansky O.~L.,  Kyuberis A.~A.,  Lodi L.,  Tennyson J.,  Ovsyannikov R.~I.,
   Zobov N.,  2017, Mon. Not. R. Astron. Soc., 466, 1363

\bibitem[\protect\citeauthoryear{Polyansky, Ovsyannikov, Kyuberis, Lodi,
  Tennyson \& Zobov}{Polyansky et~al.}{2013}]{jt550}
Polyansky O.~L.,  Ovsyannikov R.~I.,  Kyuberis A.~A.,  Lodi L.,  Tennyson J.,
   Zobov N.~F.,  2013, J. Phys. Chem. A, 117, 9633–9643

\bibitem[\protect\citeauthoryear{Polyansky, Zobov, Tennyson, Lotoski \&
  Bernath}{Polyansky et~al.}{1997}]{jt203}
Polyansky O.~L.,  Zobov N.~F.,  Tennyson J.,  Lotoski J.~A.,    Bernath P.~F.,
  1997, J. Mol. Spectrosc., 184, 35

\bibitem[\protect\citeauthoryear{Polyansky, Zobov, Viti \& Tennyson}{Polyansky
  et~al.}{1998}]{jt218}
Polyansky O.~L.,  Zobov N.~F.,  Viti S.,    Tennyson J.,  1998, J. Mol.
  Spectrosc., 189, 291

\bibitem[\protect\citeauthoryear{Polyansky, Zobov, Viti, Tennyson, Bernath \&
  Wallace}{Polyansky et~al.}{1997}]{jt205}
Polyansky O.~L.,  Zobov N.~F.,  Viti S.,  Tennyson J.,  Bernath P.~F.,
  Wallace L.,  1997, Astrophys. J., 489, L205

\bibitem[\protect\citeauthoryear{Rajpurohit, Reyle, Allard, Scholz, Homeier,
  Schultheis \& Bayo}{Rajpurohit et~al.}{2014}]{14RaReAl.H2O}
Rajpurohit A.~S.,  Reyle C.,  Allard F.,  Scholz R.~D.,  Homeier D.,
  Schultheis M.,    Bayo A.,  {2014}, Astron. Astrophys., {564}, A90

\bibitem[\protect\citeauthoryear{Rein \& Sanders}{Rein \&
  Sanders}{2010}]{10ReSa.H2O}
Rein K.~D.,  Sanders S.~T.,  {2010}, Appl. Optics, {49}, 4728

\bibitem[\protect\citeauthoryear{Rothman, Gordon, Barber, Dothe, Gamache,
  Goldman, Perevalov, Tashkun \& Tennyson}{Rothman et~al.}{2010}]{jt480}
Rothman L.~S.,  Gordon I.~E.,  Barber R.~J.,  Dothe H.,  Gamache R.~R.,
  Goldman A.,  Perevalov V.~I.,  Tashkun S.~A.,    Tennyson J.,  2010, J.
  Quant. Spectrosc. Radiat. Transf., 111, 2139

\bibitem[\protect\citeauthoryear{Rutkowski, Foltynowicz, Johansson,
  Khodabakhsh, Schmidt, Kyuberis, Zobov, Polyansky, Yurchenko \&
  Tennyson}{Rutkowski et~al.}{2018}]{jt712}
Rutkowski L.,  Foltynowicz A.,  Johansson A.~C.,  Khodabakhsh A.,  Schmidt
  F.~M.,  Kyuberis A.~A.,  Zobov N.~F.,  Polyansky O.~L.,  Yurchenko S.~N.,
  Tennyson J.,  2018, J. Quant. Spectrosc. Radiat. Transf., 205, 213

\bibitem[\protect\citeauthoryear{Ryde, Lambert, Farzone, Richter, Josselin,
  Harper, Eriksson \& Greathouse}{Ryde et~al.}{2015}]{15RyLaFa.H2O}
Ryde N.,  Lambert J.,  Farzone M.,  Richter M.~J.,  Josselin E.,  Harper G.~M.,
   Eriksson K.,    Greathouse T.~K.,  {2015}, Astron. Astrophys., {573}, A28

\bibitem[\protect\citeauthoryear{Ryde, Richter, Harper, Eriksson \&
  Lambert}{Ryde et~al.}{2006}]{06RyRiHa.H2O}
Ryde N.,  Richter M.~J.,  Harper G.~M.,  Eriksson K.,    Lambert D.~L.,
  {2006}, Astrophys. J., {645}, 652

\bibitem[\protect\citeauthoryear{Schermaul, Learner, Canas, Brault, Polyansky,
  Belmiloud, Zobov \& Tennyson}{Schermaul et~al.}{2002}]{jt285}
Schermaul R.,  Learner R. C.~M.,  Canas A. A.~D.,  Brault J.~W.,  Polyansky
  O.~L.,  Belmiloud D.,  Zobov N.~F.,    Tennyson J.,  2002, J. Mol.
  Spectrosc., 211, 169

\bibitem[\protect\citeauthoryear{Schwenke}{Schwenke}{2001}]{Schwenke2001}
Schwenke D.~W.,  2001, J. Phys. Chem. A, 105, 2352

\bibitem[\protect\citeauthoryear{Shirin, Polyansky, Zobov, Barletta \&
  Tennyson}{Shirin et~al.}{2003}]{jt308}
Shirin S.~V.,  Polyansky O.~L.,  Zobov N.~F.,  Barletta P.,    Tennyson J.,
  2003, J. Chem. Phys., 118, 2124

\bibitem[\protect\citeauthoryear{Shirin, Polyansky, Zobov, Ovsyannikov,
  {Cs\'asz\'ar} \& Tennyson}{Shirin et~al.}{2006}]{jt375}
Shirin S.~V.,  Polyansky O.~L.,  Zobov N.~F.,  Ovsyannikov R.~I.,
  {Cs\'asz\'ar} A.~G.,    Tennyson J.,  2006, J. Mol. Spectrosc., 236, 216

\bibitem[\protect\citeauthoryear{Shirin, Zobov, Ovsyannikov, Polyansky \&
  Tennyson}{Shirin et~al.}{2008}]{jt438}
Shirin S.~V.,  Zobov N.~F.,  Ovsyannikov R.~I.,  Polyansky O.~L.,    Tennyson
  J.,  2008, J. Chem. Phys., 128, 224306

\bibitem[\protect\citeauthoryear{Sonnabend, Wirtz, Schieder \&
  Bernath}{Sonnabend et~al.}{2006}]{06SoWiSc.H2O}
Sonnabend G.,  Wirtz D.,  Schieder R.,    Bernath P.,  {2006}, Solar Phys.,
  {233}, 205

\bibitem[\protect\citeauthoryear{Tennyson, Bernath, Brown, Campargue, Carleer,
  Cs\'asz\'ar, Daumont, Gamache, Hodges, Naumenko, Polyansky, Rothmam,
  Vandaele, Zobov, {Al Derzi}, F\'abri, Fazliev, Furtenbacher, Gordon, Lodi \&
  Mizus}{Tennyson et~al.}{2013}]{jt539}
Tennyson J.,  Bernath P.~F.,  Brown L.~R.,  Campargue A.,  Carleer M.~R.,
  Cs\'asz\'ar A.~G.,  Daumont L.,  Gamache R.~R.,  Hodges J.~T.,  Naumenko
  O.~V.,  Polyansky O.~L.,  Rothmam L.~S.,  Vandaele A.~C.,  Zobov N.~F.,  {Al
  Derzi} A.~R.,  F\'abri C.,  Fazliev A.~Z.,  Furtenbacher T.,  Gordon I.~E.,
  Lodi L.,    Mizus I.~I.,  2013, J. Quant. Spectrosc. Radiat. Transf., 117, 29

\bibitem[\protect\citeauthoryear{Tennyson, Bernath, Brown, Campargue, Carleer,
  Cs\'asz\'ar, Daumont, Gamache, Hodges, Naumenko, Polyansky, Rothman, Toth,
  Vandaele, Zobov, Fazliev, Furtenbacher, Gordon, Mikhailenko \&
  Voronin}{Tennyson et~al.}{2010}]{jt482}
Tennyson J.,  Bernath P.~F.,  Brown L.~R.,  Campargue A.,  Carleer M.~R.,
  Cs\'asz\'ar A.~G.,  Daumont L.,  Gamache R.~R.,  Hodges J.~T.,  Naumenko
  O.~V.,  Polyansky O.~L.,  Rothman L.~S.,  Toth R.~A.,  Vandaele A.~C.,  Zobov
  N.~F.,  Fazliev A.~Z.,  Furtenbacher T.,  Gordon I.~E.,  Mikhailenko S.~N.,
   Voronin B.~A.,  2010, J. Quant. Spectrosc. Radiat. Transf., 111, 2160

\bibitem[\protect\citeauthoryear{Tennyson, Bernath, Brown, Campargue, Carleer,
  Cs\'asz\'ar, Gamache, Hodges, Jenouvrier, Naumenko, Polyansky, Rothman, Toth,
  Vandaele, Zobov, Daumont, Fazliev, Furtenbacher, Gordon, Mikhailenko \&
  Shirin}{Tennyson et~al.}{2009}]{jt454}
Tennyson J.,  Bernath P.~F.,  Brown L.~R.,  Campargue A.,  Carleer M.~R.,
  Cs\'asz\'ar A.~G.,  Gamache R.~R.,  Hodges J.~T.,  Jenouvrier A.,  Naumenko
  O.~V.,  Polyansky O.~L.,  Rothman L.~S.,  Toth R.~A.,  Vandaele A.~C.,  Zobov
  N.~F.,  Daumont L.,  Fazliev A.~Z.,  Furtenbacher T.,  Gordon I.~E.,
  Mikhailenko S.~N.,    Shirin S.~V.,  2009, J. Quant. Spectrosc. Radiat.
  Transf., 110, 573

\bibitem[\protect\citeauthoryear{Tennyson, Bernath, Brown, Campargue,
  Cs\'asz\'ar, Daumont, Gamache, Hodges, Naumenko, Polyansky, Rothmam,
  Vandaele, Zobov, D\'enes, Fazliev, Furtenbacher, Gordon, Hu, Szidarovszky \&
  Vasilenko}{Tennyson et~al.}{2014b}]{jt576}
Tennyson J.,  Bernath P.~F.,  Brown L.~R.,  Campargue A.,  Cs\'asz\'ar A.~G.,
  Daumont L.,  Gamache R.~R.,  Hodges J.~T.,  Naumenko O.~V.,  Polyansky O.~L.,
   Rothmam L.~S.,  Vandaele A.~C.,  Zobov N.~F.,  D\'enes N.,  Fazliev A.~Z.,
  Furtenbacher T.,  Gordon I.~E.,  Hu S.-M.,  Szidarovszky T.,    Vasilenko
  I.~A.,  2014b, J. Quant. Spectrosc. Radiat. Transf., 142, 93

\bibitem[\protect\citeauthoryear{Tennyson, Bernath, Brown, Campargue,
  Cs\'asz\'ar, Daumont, Gamache, Hodges, Naumenko, Polyansky, Rothman, Vandaele
  \& Zobov}{Tennyson et~al.}{2014a}]{jt562}
Tennyson J.,  Bernath P.~F.,  Brown L.~R.,  Campargue A.,  Cs\'asz\'ar A.~G.,
  Daumont L.,  Gamache R.~R.,  Hodges J.~T.,  Naumenko O.~V.,  Polyansky O.~L.,
   Rothman L.~S.,  Vandaele A.~C.,    Zobov N.~F.,  2014a, Pure Appl. Chem., 86,
  71

\bibitem[\protect\citeauthoryear{Tennyson, Hill \& Yurchenko}{Tennyson
  et~al.}{2013}]{jt548}
Tennyson J.,  Hill C.,    Yurchenko S.~N.,  2013, in 6$^{th}$ international
  conference on atomic and molecular data and their applications ICAMDATA-2012
  Vol.~1545 of AIP Conference Proceedings, {Data structures for ExoMol:
  Molecular line lists for exoplanet and other atmospheres}.
AIP, New York, pp 186--195

\bibitem[\protect\citeauthoryear{Tennyson, Kostin, Barletta, Harris, Polyansky,
  Ramanlal \& Zobov}{Tennyson et~al.}{2004}]{jt338}
Tennyson J.,  Kostin M.~A.,  Barletta P.,  Harris G.~J.,  Polyansky O.~L.,
  Ramanlal J.,    Zobov N.~F.,  2004, Comput. Phys. Commun., 163, 85

\bibitem[\protect\citeauthoryear{Tennyson, Miller \& {Le Sueur}}{Tennyson
  et~al.}{1993}]{jt128}
Tennyson J.,  Miller S.,    {Le Sueur} C.~R.,  1993, Comput. Phys. Commun., 75,
  339

\bibitem[\protect\citeauthoryear{Tennyson \& Yurchenko}{Tennyson \&
  Yurchenko}{2012}]{jt528}
Tennyson J.,  Yurchenko S.~N.,  2012, Mon. Not. R. Astron. Soc., 425, 21

\bibitem[\protect\citeauthoryear{Tennyson \& Yurchenko}{Tennyson \&
  Yurchenko}{2017}]{jt626}
Tennyson J.,  Yurchenko S.~N.,  2017, Intern. J. Quantum Chem., 117, 92

\bibitem[\protect\citeauthoryear{Tennyson \& Yurchenko}{Tennyson \&
  Yurchenko}{2018}]{jt731}
Tennyson J.,  Yurchenko S.~N.,  2018, Atoms, 6, 26

\bibitem[\protect\citeauthoryear{Tennyson, Yurchenko, Al-Refaie, Barton, Chubb,
  Coles, Diamantopoulou, Gorman, Hill, Lam, Lodi, McKemmish, Na, Owens,
  Polyansky, Rivlin, Sousa-Silva, Underwood, Yachmenev \& Zak}{Tennyson
  et~al.}{2016}]{jt631}
Tennyson J.,  Yurchenko S.~N.,  Al-Refaie A.~F.,  Barton E.~J.,  Chubb K.~L.,
  Coles P.~A.,  Diamantopoulou S.,  Gorman M.~N.,  Hill C.,  Lam A.~Z.,  Lodi
  L.,  McKemmish L.~K.,  Na Y.,  Owens A.,  Polyansky O.~L.,  Rivlin T.,
  Sousa-Silva C.,  Underwood D.~S.,  Yachmenev A.,    Zak E.,  2016, J. Mol.
  Spectrosc., 327, 73

\bibitem[\protect\citeauthoryear{Tinetti, Vidal-Madjar, Liang, Beaulieu, Yung,
  Carey, Barber, Tennyson, Ribas, Allard, Ballester, Sing \& Selsis}{Tinetti
  et~al.}{2007}]{jt400}
Tinetti G.,  Vidal-Madjar A.,  Liang M.-C.,  Beaulieu J.-P.,  Yung Y.,  Carey
  S.,  Barber R.~J.,  Tennyson J.,  Ribas I.,  Allard N.,  Ballester G.~E.,
  Sing D.~K.,    Selsis F.,  2007, Nature, 448, 169

\bibitem[\protect\citeauthoryear{Tolchenov, Naumenko, Zobov, Shirin, Polyansky,
  Tennyson, Carleer, Coheur, Fally, Jenouvrier \& Vandaele}{Tolchenov
  et~al.}{2005}]{jt360}
Tolchenov R.~N.,  Naumenko O.,  Zobov N.~F.,  Shirin S.~V.,  Polyansky O.~L.,
  Tennyson J.,  Carleer M.,  Coheur P.-F.,  Fally S.,  Jenouvrier A.,
  Vandaele A.~C.,  2005, J. Mol. Spectrosc., 233, 68

\bibitem[\protect\citeauthoryear{Tsuji}{Tsuji}{2001}]{01Tsujix.H2O}
Tsuji T.,  {2001}, Astron. Astrophys., {376}, L1

\bibitem[\protect\citeauthoryear{Underwood, Tennyson, Yurchenko, Huang,
  Schwenke, Lee, Clausen \& Fateev}{Underwood et~al.}{2016}]{jt635}
Underwood D.~S.,  Tennyson J.,  Yurchenko S.~N.,  Huang X.,  Schwenke D.~W.,
  Lee T.~J.,  Clausen S.,    Fateev A.,  2016, Mon. Not. R. Astron. Soc., 459,
  3890

\bibitem[\protect\citeauthoryear{Varandas}{Varandas}{1996}]{Varandas96}
Varandas A. J.~C.,  {1996}, J. Chem. Phys., pp 3524--3531

\bibitem[\protect\citeauthoryear{Vidler \& Tennyson}{Vidler \&
  Tennyson}{2000}]{jt263}
Vidler M.,  Tennyson J.,  2000, J. Chem. Phys., 113, 9766

\bibitem[\protect\citeauthoryear{Viti, Tennyson \& Polyansky}{Viti
  et~al.}{1997}]{jt197}
Viti S.,  Tennyson J.,    Polyansky O.~L.,  1997, Mon. Not. R. Astron. Soc.,
  287, 79

\bibitem[\protect\citeauthoryear{Voronin, Tennyson, Tolchenov, Lugovskoy \&
  Yurchenko}{Voronin et~al.}{2010}]{jt469}
Voronin B.~A.,  Tennyson J.,  Tolchenov R.~N.,  Lugovskoy A.~A.,    Yurchenko
  S.~N.,  2010, Mon. Not. R. Astron. Soc., 402, 492

\bibitem[\protect\citeauthoryear{Yurchenko, Al-Refaie \& Tennyson}{Yurchenko
  et~al.}{2018}]{jt708}
Yurchenko S.~N.,  Al-Refaie A.~F.,    Tennyson J.,  2018, Astron. Astrophys.,
  614, A131

\bibitem[\protect\citeauthoryear{Yurchenko, Carvajal, Jensen, Herregodts \&
  Huet}{Yurchenko et~al.}{2003}]{03YuCaJe.PH3}
Yurchenko S.~N.,  Carvajal M.,  Jensen P.,  Herregodts F.,    Huet T.~R.,
  2003, Chem. Phys., 290, 59

\bibitem[\protect\citeauthoryear{Zobov, Shirin, Lodi, Silva, Tennyson,
  Cs\'asz\'ar \& Polyansky}{Zobov et~al.}{2011}]{jt494}
Zobov N.~F.,  Shirin S.~V.,  Lodi L.,  Silva B.~C.,  Tennyson J.,  Cs\'asz\'ar
  A.~G.,    Polyansky O.~L.,  2011, Chem. Phys. Lett., 507, 48

\end{thebibliography}

\label{lastpage}
\end{document}